\documentclass[aps,prb,twocolumn,superscriptaddress,showpacs,amsmath,amssymb,longbibliography]{revtex4-1}
\usepackage{graphicx,bbm}
\usepackage{bm}
\usepackage{color}
\usepackage{hyperref}

\newcommand{\un}[1]{{\underline{#1}}}

\def\one{\mathbbm{1}}

\def\one{\mathbbm{1}}

\begin{document}
\title{Approximate conservation laws in perturbed integrable lattice models}
\author {Marcin Mierzejewski}
\affiliation{Institute of Physics, University of Silesia, 40-007 Katowice, Poland}
\author{Toma\v z Prosen}
\affiliation{Faculty of Mathematics and Physics, University of Ljubljana, SI-1000 Ljubljana, Slovenia }
\author{Peter Prelov\v sek}
\affiliation{Faculty of Mathematics and Physics, University of Ljubljana, SI-1000 Ljubljana, Slovenia }
\affiliation{J. Stefan Institute, SI-1000 Ljubljana, Slovenia }
\begin{abstract}
We develop a numerical algorithm for identifying approximately conserved quantities in models perturbed away from integrability.
In the long--time regime, these quantities fully determine correlation functions of local observables.  
Applying the algorithm to the perturbed XXZ model we find that the main effect of perturbation consists in expanding the support of
conserved quantities. This expansion follows quadratic dependence on the strength of perturbation. The latter result together with
correlation functions of conserved quantities obtained from the memory function analysis confirm feasibility of the perturbation theory.
\end{abstract}
% 72.10.-d Theory of electronic transport; scattering mechanisms
% 71.27.+a Strongly correlated electron systems; heavy fermions
% 72.10.Bg General formulation of transport theory
% 75.10.Pq Spin chain models
% 75.10.Jm Quantized spin models, including quantum spin frustratio
% 05.60.Gg Quantum transport
% 05.70.Ln Nonequilibrium and irreversible thermodynamics

\maketitle

\section{Introduction} 
%Recent advances in physics of ultracold atoms together with modern computational techniques gave new insight into the dynamics of closed many-–body quantum systems \cite{}. 
A considerable interest has recently been attracted by integrable quantum models which, in contrast to generic systems, 
have macroscopic number of local conserved--quantities (CQ).  Due to their presence, the isolated integrable systems don't thermalize \cite{polkrev,Goldstein2006,Linden2009,Riera2012,sirker2014,konstantinidis2015} but instead relax towards non-thermal steady states.\cite{Hershfield1993,Natan2006,santos2011,our2013} It has been suggested that such unusual steady states 
are fully specified by local \cite{gge,Eckstein2012,Cassidy2011,Gogolin2011,essler2014} as well as quasilocal CQ. Inclusion of the latter ones is necessary 
at least in several systems which cannot be mapped to noninteracting particles.\cite{rigol2014,nongge1,nongge2,nongge3,tomaz_quasilocal11,tomaz_quasilocal13,tomaz_quasilocal14,affleck14,mierzejewski2015,ilievski}
The research on integrable systems has been motivated not only by such fundamental problems like that concerning mechanisms of thermalization/relaxation. 
The integrable systems are interesting also because they show dissipationless (ballistic)  energy, spin and charge transport \cite{tetelman,grabowski,zotos1996,zotos1997, zotos1999,zotos_unpub,benz,review2007,shastry, herbrych2011, Marko2011,Sirker2009,robin2013,Tomaz2011,my3,vidmar2013,crivelli2014,mendoza2015} which might be important for the future applications.

However, real systems are never perfect and their description in terms of integrable models should be viewed, at best,  as reasonable approximations. 
Therefore, it is important to understand the properties of systems which are weakly perturbed away from integrability.\cite{rosch2006,rosch2007,rosch1015,zotos2004,olshanii2015,essler2014int,huang2013,Eckstein2012,olshanii2012,PRL98,PRE99,PRE02}
In the case of classical mechanics, relevant formalism has been developed for more than fifty years,\cite{FPU,KAM}
whereas for quantum systems such understanding is still missing or, at least, remains largely incomplete. 
It is not evident to what extend the  breaking of integrability may be described within a single universal picture and which properties are specific for particular model and/or perturbation.   It is also not quite clear  which hallmarks of integrability disappear abruptly and which decay smoothly when the perturbation gradually increases. 

Recent studies allow  to formulate several general expectations. 
At least at the infinite time--scale and for sufficiently strong perturbation the  ballistic transport should be replaced by large but finite conductivity.\cite{zotos2004,rosch2006,rosch2007,huang2013}  However, after turning on the perturbation, the system should evolve towards a
 quasi-steady state (prethermalization  \cite{Eckstein2012}) which is analogous to the generalized Gibbs ensembles.  \cite{essler2014int}
Hence, at least for a finite time window $[0,\tau]$ and for sufficiently weak perturbation, the system maintains the main property of integrable systems, i.e., the existence
of local or quasilocal operators  which for perturbed systems are conserved only approximately.  These quantities are not identical with CQ of the integrable 
parent model \cite{rosch2006,essler2014int,olshanii2015} but rather they are modified by the perturbation. Finally, since quasilocal CQ play important role for strictly integrable models  \cite{nongge1,nongge2,nongge3,tomaz_quasilocal11,tomaz_quasilocal13,tomaz_quasilocal14,affleck14,mierzejewski2015,ilievski} they should also be included in the studies on perturbed systems.

In this paper we develop an algorithm  which captures/verifies all these properties of the perturbed integrable systems. 
The approach yields approximately conserved quantities (ACQ)  which completely determine the long--time correlation functions of all local operators supported on assumed subsystem. 
The algorithm   captures cases ranging from strict integrability with local and quasilocal CQ to generic systems where CQ are generally nonlocal
linear combinations of projections on eigenstates of the Hamiltonian  $|n\rangle \langle n|$.
In the first part of this paper we show the general approach, while in the second part we apply it to 
the perturbed anisotropic Heisenberg (XXZ) model.  We also find that the ACQ at weak perturbation can  be described
as quantities decaying exponentially in time with the characteristic rate depending quadratically on the perturbation
strength, consistent with specific findings of Ref.~\onlinecite{rosch2007}.
This result as well as the memory function analysis confirm non-singular behavior at weak perturbation strength and the feasibility of the perturbation theory for a generic class of integrability-breaking perturbations.

%The technical details are presented in the supplementary material.
 
\section{General method}  
We study a one-dimensional tight--binding Hamiltonian $H$ on a lattice of $L$ sites 
with periodic boundary conditions. We consider the space ${\cal A}_L$ of local, extensive, 
translationally--invariant observables with the scalar product (see Appendix A)
\begin{equation}
(A|B) =\frac{1}{L} \langle A^{\dagger} B \rangle
=
\frac{1}{ L} \sum_{mn} A^{*}_{mn} B_{mn} p(E_n).
\label{ip1}
\end{equation}   
where $A_{mn}=\langle m | A  | n \rangle $, $H  | n \rangle = E_n  | n \rangle $ and the weights are assumed to satisfy $\sum_n p(E_n)=1$ and $p(E_n) > 0$ for all $n$.
The latter assumption excludes 
the zero--temperature case but, at least in finite systems, accounts for the thermal states $p(E_n) \propto \exp(-\beta E_n)$, where $\beta$ is the inverse temperature.

We have recently developed a procedure for identifying a complete set of local and quasilocal CQ in integrable lattice systems.\cite{mierzejewski2015}
The main step is to construct  scalar products of all time--averaged operators $(\bar{A}|\bar{B})$, where 
\begin{equation}
\bar{A}= \lim_{\tau \to \infty}  \frac{1}{\tau} \int_0^{\tau} dt A(t) = \sum_{E_m=E_n}  A_{mn}  |m \rangle \langle n|. 
\end{equation}   
The identity $(\bar{A}|\bar{B})=(\bar{A}|B)$ is essential since it allows to distinguish  between local, quasilocal and  generic nonlocal CQ.  In order to determine ACQ in the perturbed system one should consider a finite time--window  $[0,\tau]$, however, a simple omission of the limit $\tau \rightarrow \infty$ 
violates the latter essential relation. Therefore, we define an effective operator time-average with a time-scale $\tau$ as  
\begin{equation}
\bar{A}^{\tau} =   \int_{-\infty}^\infty {\rm d} t\; A(t) \frac{\sin(t/\tau)}{\pi t}
\end{equation}
which in spectral representation amounts to cutting off quickly oscillating (in time) matrix elements
\begin{eqnarray}
\bar{A}^{\tau} & = & \sum_{mn}  \theta \left(\frac{1}{\tau}-|E_n-E_m| \right)  A_{mn}  |m \rangle \langle n|. \label{ta}
 \end{eqnarray}
It is quite obvious that the truncated operators  are approximately conserved at the time--scale  $t \ll \tau$. 
Since $\theta^2(x)=\theta(x)$, this simplified time--averaging maintains the property  $(\bar{A}^{\tau} |\bar{B}^{\tau} )=(\bar{A}^{\tau}|B)$. 
Moreover, it becomes identical with the actual time--averaging over an infinite time--window,  $\lim_{\tau \to \infty} \bar{A}^{\tau} =  \bar{A} $, whereas 
for finite $\tau$ it is related with the low-frequency spectrum of standard correlation functions:
\begin{equation}
(\bar{A}^{\tau}|\bar{B}^{\tau})= \lim_{\varepsilon\searrow 0}\int_{-\frac{1}{\tau}}^{\frac{1}{\tau}} {\rm d} \omega \;\; \frac{1}{2 \pi} \int_{-\infty}^{\infty} {\rm d} t e^{i \omega t-|t|\varepsilon} \frac{\langle A^{\dagger} B(t) \rangle}{L}. 
\label{corfun}
\end{equation}

%In order to single out the relevant local and quasilocal operators we apply the approach derived in Ref. \cite{mierzejewski2015}. 

Since physically interesting observables are usually supported on few sites only, we define a subspace ${\cal B}^{M}_L$ of ${\cal A}_L$ which contains operators supported on up to $M$ consecutive lattice sites.  The choice of interesting operators determines $M$. 
  We introduce also  the basis of ${\cal B}^M_L$  composed of operators $O_s$ which are orthonormal $(O_s|O_{s'})=\delta_{s,s'}$. 
After constructing $O_s$  we solve the eigenproblem for the matrix $K^\tau_{ss'}= (\bar{O}^{\tau}_s|\bar{O}^{\tau}_{s'})$
\begin{equation}
\sum_{s,s'} U^{\dagger}_{ls} K^\tau_{ss'}  U_{s'l'} = \delta_{ll'} \lambda_{l},\quad  U U^{\dagger}=U^{\dagger}U=1, \label{eig} \\
\end{equation}
which generates orthogonal set of  ACQ consisting of  $Q_l  =  \sum_{s} U_{sl} \bar{O}^{\tau}_s $.  
Generally, the truncation (\ref{ta}) modifies the support of operators and transforms local operators $O_s$ into quasilocal ones $\bar{O}^{\tau}_s$.
Therefore, we split $Q_l$ into two orthogonal components $Q_l=Q^M_l+Q^{\perp}_l$ such that $Q^M_l=\sum_s  (O_s|Q_l) O_s \in {\cal B}^M_L$ while $(Q^{\perp}_l|O_s)=0$ for all $s$.
The eigenvalues $\lambda_l$ obtained from Eq. (\ref{eig})  bear important information on the support of $Q_l$ (see Appendix B),
\begin{equation}
 ||Q_l||^2 =  (Q_l|Q_l)=\lambda_l,\quad\quad
 \frac{||Q^M_l||^2}{||Q_l||^2}=\lambda_l.  \label{sup}
\end{equation}
Carrying out the finite size scaling ($L \rightarrow \infty$) of $\lambda_l$  we distinguish between local ACQ when $\lambda_l=1$ for sufficiently large $M$, quasilocal ACQ 
when $0 < \lambda_l <1$ for any $M$, and generic nonlocal ACQ when $\lambda_l \rightarrow 0$. One can also show (see Appendix C) that the correlation function of {\em arbitrary local} observables 
$A,B \in {\cal B}^{M}_L$ is completely determined by their projections on $Q_l$,
 \begin{eqnarray}
(\bar{A}^{\tau}|\bar{B}^{\tau}) & = & \sum_{l} \frac{(A|Q_l)(Q_l|B)}{(Q_l|Q_l)}  \label{compg} \\ 
  &=&   \sum_{l} \lambda_l \frac{(A|Q^M_l) (Q^M_l|B)}{(Q^M_l|Q^M_l)}, \label{compl}
\end{eqnarray}
where the latter equation comes from identity $(A|Q^{\perp}_l)=0$.
 Choosing $A=B$ as a current operator and taking the limit $\tau \rightarrow \infty$ we recognize that Eq. (\ref{compg}) becomes the saturated Mazur bound \cite{mazur,zotos1997} for the charge/spin stiffness.  Note that Eq. (\ref{compg}) involves normalized operators $Q_l/||Q_l||$. 
Therefore, the key point is to follow how the supports of ACQ (and not their norms) depend on the time-scale $\tau$ and the strength of perturbation.\cite{sirker2014}
Our approach gives complete set of ACQ, which are sorted from the most relevant local operators with $\lambda_l=1$  to the least relevant ACQ with the smallest $\lambda_l$.

\section{Perturbed Anisotropic Heisenberg model} Next, we apply this approach to the extended XXZ model
\begin{eqnarray}
H &=& J \sum_{j=1}^L  \left[ \frac{1}{2} (S_j^+ S_{j+1}^- + S_j^- S_{j+1}^+)  + \Delta S_j^z S_{j+1}^z \right]+\alpha H'  \nonumber \\
H'&=& J \sum_{j=1}^L S_j^z S_{j+2}^z  , \label{H}
\end{eqnarray}
where $S_j^{\pm,z}$ are spin-$1/2$ operators and the integrability is broken by the last term when  $\alpha \ne 0$. We take $J=1$ as the energy unit.
For concreteness, we study the infinite--temperature limit  [$p(E_n)= {\rm const}$]  when Eq. (\ref{ip1})  becomes the Hilbert-Schmidt scalar product. 
Then, the orthonormal basis of ${\cal B}^M_L$ is composed of operators \cite{mierzejewski2015}
\begin{equation}
O_{\un{s}} = \sum_j \sigma^{s_1}_j \sigma^{s_2}_{j+1} \cdots \sigma^{s_m}_{j+m-1},\quad \quad m=2,...,M \label{os}
\end{equation}
where $\sigma_j^z\equiv2S_j^z,\sigma_j^\pm\equiv\sqrt{2}S_j^\pm,\sigma^0_j\equiv\one,\un{s}=(s_1,\ldots,s_m)$,  
$s_j\in\{+,-,z,0\}$ while $s_{1,m}\in\{+,-,z\}$.
We introduce symbols ``R'' and ``I'' to distinguishes between real ($O_{\un{s}}+O^{\dagger}_{\un{s}}$) and imaginary ($iO_{\un{s}}-iO^{\dagger}_{\un{s}}$)  combinations of basis operators, respectively. We use also letters ``E'' and ``O'' to distinguishes between operators which respectively are odd and even under the spin--flip transformation. Since the Hamiltonian is invariant under spin--flip and time--reversal transformation, we separately study four orthogonal sectors of operators denoted as  RE (includes, e.g., the Hamiltonian), IE (includes, e.g., the energy current), IO (includes, e.g., the spin current), and RO. In the integrable parent model, the local and quasilocal CQ exist in all four sectors provided $\Delta < 1$.\cite{mierzejewski2015} In order not to exclude any symmetry sector from our considerations, we take $\Delta=\frac{1}{2}$. We restrict also the Hilbert space to the states with $S^{z}_{tot}=0$. 

\begin{figure} 
\includegraphics[width=0.48\textwidth]{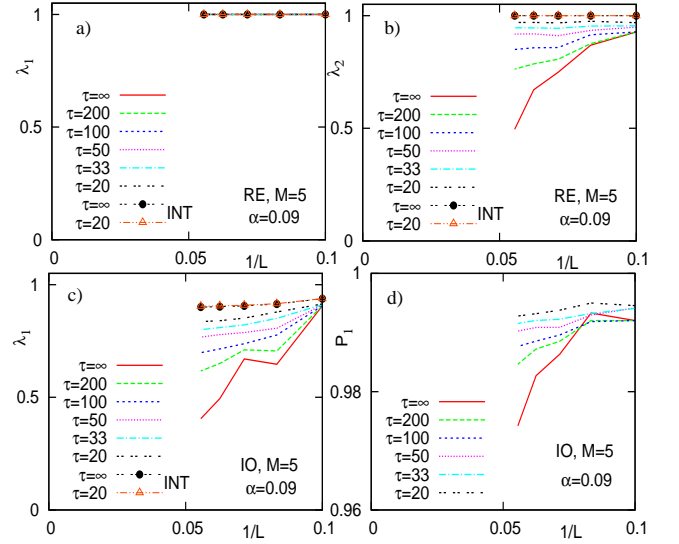}
\caption{Size-dependence of the leading eigenvalues $\lambda_l$ [Eq. (\ref{eig})] and projections $P_l$ [Eq. (\ref{proj})] for various time-windows $[0,\tau]$.
The symmetry sector (RE or IO) and other parameters are shown in each panel. Lines with points  in panels a,b and c show results for the parent integrable model.
Note different scale in  d. 
}
\label{fig1}
\end{figure}

As follows from Eq, (\ref{compl}), breaking of integrability affects the correlation functions by either changing 
the support of $Q_l$ (parameterized by $\lambda_l$) or by changing the projected operators, $Q^M_l$. 
In order to quantify the latter changes we have calculated the projections   
\begin{equation}
P_l=\frac{(Q^M_l | Q^M_{0,l})}{||Q^{M}_l||\;||Q^{M}_{0,l}||},
\label{proj}
\end{equation}
where $Q_{0,l}$ are CQ obtained for the integrable parent model.  
Figure \ref{fig1} shows size--dependence of $\lambda_l$ and $P_l$ for various $\tau$.  
We observe that the perturbation strongly reduces $\lambda_l$ (except for $\lambda_1$ in RE sector discussed below), whereas the projected operators $Q^M_l$ do not change significantly.
Therefore, we conclude that the  main effect of perturbation consists in expanding the support of ACQ. 
%This holds true at least for operators supported on $M \le 5$ sites which are accessible to our numerical facilities.  
%However, it is clear that $\lambda_l$ should gradually grow when $M$ increases. Eventually for much larger $M$, the advantage from using projected ACQ  [$Q^M_l$ in Eq. (\ref{compl})] instead of bare ACQ [$Q_l$ in Eq.(\ref{compg})] may fade out. 

 From now on we focus on the support of ACQ as parameterized by $\lambda_l$ [see Eq. (\ref{sup})]. The leading eigenvalues in the parent integrable model (lines with points in Fig. \ref{fig1}) 
are independent of the time--scale $\tau$ indicating that the corresponding $Q_{0,l}$ are strictly conserved. It holds true both
for local CQ with $\lambda_l=1$ (Figs. \ref{fig1}a and \ref{fig1}b) as well as quasilocal CQ with  $0<\lambda_l<1$ (Fig. \ref{fig1}c). 
For the perturbed system, the only strictly conserved quantity is $Q_1$ in the RE sector, which actually represents the Hamiltonian.
All other $Q_l$ are quasilocal for finite $\tau$.  Their supports visibly depend on the time--scale even for quite large $\tau$, hence they are conserved only approximately.
We have verified these conclusions also for other eigenvalues,  symmetry sectors, supports $M=3,4,5$ and perturbations $\alpha/\Delta=\frac{1}{8},\frac{1}{6},\frac{1}{4}$.

\begin{figure}
\includegraphics[width=0.48\textwidth]{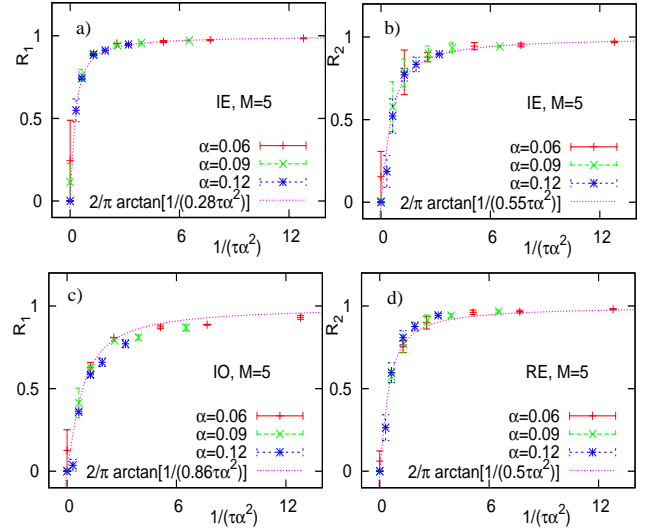} 
\caption{Extrapolated  and normalized eigenvalues $\lambda_l$ [see Eq. \ref{r}] as a function of $1/\tau \alpha^2$. The vertical sizes of symbols show the differences between
linear extrapolation from  $L=12,14,16,18$ and $L=14,16,18$. }
\label{fig2}
\end{figure}

The most relevant and challenging problem is to establish $\lambda_l$ for large $\tau$ and small $\alpha$.  
It is also important that the finite size scaling $L \rightarrow \infty$  precedes the limit $\tau \rightarrow \infty$.\cite{sirker2014}
In Fig. \ref{fig2} we plot $\lambda_l$ linearly extrapolated to $1/L \rightarrow 0$ and normalized to results for integrable parent model.
We clearly see that the dependence of extrapolated $\lambda_l$ on $\tau$ and $\alpha$ is universal
\begin{eqnarray}
R_l(\tau,\alpha) &=&\frac{\lambda_l(L\rightarrow \infty,\tau,\alpha)}{\lambda_l(L\rightarrow \infty,\tau \rightarrow \infty,\alpha=0)} 
\simeq  \tilde{R}_l( \tau \alpha^2), \label{r}
 \nonumber \\
\end{eqnarray} 
and can be well approximated by 
\begin{eqnarray}
R_l(\tau,\alpha) \simeq \frac{2}{\pi} \arctan\left( \frac{1}{\tau \alpha^2 \gamma_l } \right). \label{ra}
\end{eqnarray}  
We have found the same type of behavior for other eigenvalues (excluding the Hamiltonian),  symmetry sectors and accessible supports $M=3,4,5$ (not shown).
  
Our main result for the Heisenberg model at infinite temperature  [Eqs. (\ref{r}) and $(\ref{ra})$]  can be best explained 
in the formalism of the 
%the dynamical auto-correction functions of  local and quasilocal CQ (conserved in integrable system $\alpha=0$) and the corresponding 
memory functions. We apply the projection procedure
according to Mori\cite{mori65} (see also Refs. \onlinecite{rosch2006,rosch2007,rosch1015}), to analyze the relaxation function for CQ of the integrable parent model, $Q_{0,l}$
\begin{eqnarray}
\Phi_l(\omega)&=&  (Q_{0,l}|\frac{1}{{\cal L}-\omega} | Q_{0,l}) 
%&=& \frac{\chi_{l}(\omega)- ||Q_{0,l}||^2}{\omega}
= - \frac{ ||Q_{0,l}||^2 }{\omega+M(\omega)} , \nonumber \\
M(\omega) &=& ( \bar P {\cal L} Q_{0,l}| \frac{1}{\bar P {\cal L} \bar P -\omega} |  \bar P {\cal L} Q_{0,l}), \label{mem}
\end{eqnarray}
where ${\cal L} A=[H,A]$ and $\bar P$ is the projection onto the operator  space orthogonal to $Q_{0,l}$.
%i.e. $P=|Q_{0,l}) ||Q_{0,l}||^{-2} (Q_{0,l}|$ , and the dynamical susceptibility at intfinite temperature
%$\chi_{l}(z)= \lim_{\beta \rightarrow 0} (i/\beta)  \int_0^\infty dt e^{izt} 
%\langle [Q_{0,l}^\dagger(t), Q_{0,l}]\rangle dt$.

When the formalism is applied to the perturbed integrable system (\ref{H}) with $\alpha \ll 1$, it follows directly from 
Eq.~(\ref{mem}) that 
\begin{equation}
{\cal L} Q_{0,l}=\alpha [H^\prime,Q_{0,l}] \propto \alpha, 
\end{equation}
so that $M(\omega) = \alpha^2 {\cal M}(\omega)$. It is plausible, but 
by no means obvious that the imaginary part of the  memory function, ${\cal M}''(\omega)$, is almost constant for small $|\omega|$. 
However, if the latter is true then the dynamical relaxation reduces to 
\begin{equation}
\Phi_l(\omega) \simeq -  \frac{||Q_{0,l}||^2}{\omega + i \alpha^2 {\cal M}''(\omega )} \simeq
- \frac{||Q_{0,l}||^2}{\omega + i \alpha^2 \gamma_l}. \label{lorentz} 
\end{equation}

We end up with a Lorentzian form which explains the specific dependence of eigenvalues $\lambda_l$ on the time-scale $\tau$ and perturbation $\alpha$  in Eq. (\ref{ra}). 
Namely, using Eqs. (\ref{corfun}) and (\ref{lorentz}) we find  
\begin{equation}
 \int_{-\frac{1}{\tau}}^{\frac{1}{\tau}} {\rm d} \omega \; {\rm Im} \left[\Phi_l(\omega) \right] \propto  \arctan\left( \frac{1}{\tau \alpha^2 \gamma_l } \right).
\label{intfi}
\end{equation}  

The memory function analysis can be  easily generalized also to a quantity $A$ which is not CQ but has substantial
 overlap with  the conserved quantity, 
\begin{equation}
\frac{|(A| Q_{0,l})|}{||A||\ ||Q_{0,l}||} \sim o(1).
\end{equation}
In such case the numerator  in Eq.(\ref{lorentz})
should be renormalized becoming the Drude weight (dissipationless part) of the considered operator $A$     
\begin{equation}
\Phi_l(\omega) \simeq -  \frac{||Q_{0,l}||^2 |(A| Q_{0,l})|^2}{\omega + i \alpha^2 {\cal M}''(\omega )} \simeq 
- \frac{||Q_{0,l}||^2 |(A| Q_{0,l})|^2}{\omega + i \alpha^2 \gamma_l}. \label{lorentzss}   
\end{equation}
The latter equation is valid only for weak enough perturbation $\alpha \ll 1$ and in the low $\omega$ regime. 

For the numerical calculations of the memory function  we employ the microcanonical Lanczos method,\cite{mclm} well adapted for the studies of dynamics 
at $\beta \to 0$ where we can evaluate spin systems with up to $L=32$ sites. The important
parameter is the number of Lanczos steps $N_L \leq 20000$ which determines the 
$\omega$ resolution of the method $\delta \omega \sim \delta E/ N_L$ where  $\delta E$ is the energy span
of the $L$-site spectrum, so that we reach $\delta \omega \sim 10^{-3}$. 

\begin{figure}
\includegraphics[width=0.48\textwidth]{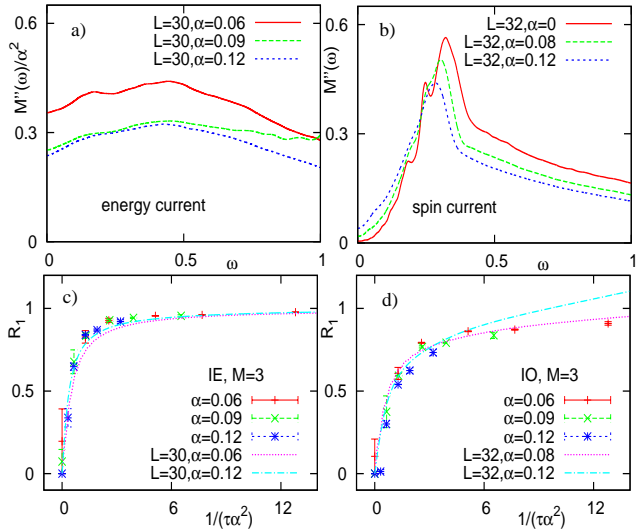}
\caption{Panels a and b show normalized $M''/\alpha^2$ for the energy current $j_E$, and $M''$ for the spin
current $j_s$, respectively.
Points in panels c and d show the same as in Fig. 2 but for $M=3$ (note that $j_E,j_s \in B^{M=3}_L$). 
Here, lines show left hand side of Eq. (\ref{intfi})  for $j_E$ (in c) and $j_s$ (in d). 
The latter quantities have been normalized to the results for integrable parent model.}
\label{fig3}
\end{figure}

We have carried out numerical calculations at $\beta \rightarrow 0$  for $Q_{0,1}$ in the IE sector being the 
energy current $j_E$ in the unperturbed parent model and for the spin current $j_s$ which has large projection on 
quasilocal $Q_{0,1}$ in the IO sector. 
From numerically obtained $\Phi^{\prime\prime}_{1}(\omega)$ we have extracted the relevant $M(\omega)$ via Eq.~(\ref{mem}).
Results presented in Fig. \ref{fig3}a confirm that $M^{\prime\prime}(\omega)/\alpha^2$ for $j_E$ is indeed very broad, featureless 
in a wide range $\omega\in[0,\omega_0]$ $\omega_0>1$, and (almost) independent of $\alpha$. 
On the other hand, $\Phi_1(\omega)$ for $j_s$ has a nonzero  $M^{\prime\prime}(\omega)$ even for integrable $\alpha=0$ case, since $j_s$ is not conserved.  
Still, in the regime $\omega \ll 1$, $M^{\prime\prime}(\omega)$ as well $\Phi_l(\omega)$ follow the scaling as
given by Eq.~(\ref{lorentz}).
% and the result presented in Fig.~3d (where $\alpha=0.12$ case already reflects some incoherent part of $M^{\prime\prime}(\omega)$).  
Note that for the Lorentzian assumption in Eq.~(\ref{lorentz}) it is enough that it holds for 
$\omega  \alt \alpha^2 \gamma_1$. Most importantly, Figs. \ref{fig3}c and \ref{fig3}d show convincing quantitative agreement between the result obtained from our general approach and 
the formalism of the memory functions. Since the latter results have been obtained for much larger  systems (but for two observables only)  they can also  serve a test  of the finite--size scaling of $\lambda_l$. 
 
%Then, Eq.~(\ref{lorentz}), is valid only for weak enough perturbation $\alpha \ll 1$ and in the low $\omega$ regime provided that the numerator is renormalized to 
%$||Q_{0,l}||^2 \rightarrow ||Q_{0,l}||^2 |(A| Q_{0,l})|^2$ representing the Drude weight (dissipationless part) for the considered $A$.  

%In order to test the analytical considerations above
%To test the feasibility of analysis above  we evaluate numerically  the dynamical relaxation $\Phi_l(\omega)$
%using the micro canonical Lanczos method (MCLM) \cite{supp} for two  quantities of particular interest within the 
%XXZ model, Eq.~(\ref{H}). Namely, 

\section{Discussion and Conclusions} We have proposed a general algorithm for a construction and quantitative description of a full set of almost conserved quasi-local operators of weakly non-integrable lattice systems.
The method has been implemented in the generic case of a XXZ model perturbed away from integrability by the 2nd nearest-neighbor interaction of strength $\alpha$.
We have obtained 
a set of orthogonal approximately conserved operators, $Q_l$. For an assumed 
time-scale $\tau$, these quantities completely determine the correlation functions of all local observables supported on several ($M$) lattice sites. 
$Q_l$ smoothly depend on $\tau$ and $\alpha$, and for the limiting case $\tau \rightarrow \infty$, $\alpha \rightarrow 0$ 
coincide with strictly conserved (local or quasilocal) quantities of integrable parent model.  
We have shown that the perturbation influences the correlation functions mostly 
by expanding the supports of $Q_l$. In our approach this effect is parameterized by eigenvalues $\lambda_l$ decreasing from $\lambda_l=1$ (for local operators) down to $\lambda_l=0$
(for generic nonlocal operators). 
We have found a scaling $\lambda_l/\lambda_l^{\alpha=0} \propto \arctan\left[1/(\tau \alpha^2 \gamma_l)\right]$ universal for all $Q_l$ except the Hamiltonian. 
For $\alpha \ne 0$, the Hamiltonian remains the only strictly local or strictly conserved quantity ($\lambda_1=1$ independently of $\tau$) while all other
$Q_l$ become quasilocal. We have found the same scaling also for a system perturbed by 4th nearest-neighbor interaction (not shown). 
This scaling seems to be typical,  however, one cannot exclude that it breaks down for other specially tuned perturbations (see, e.g., \onlinecite{rosch2006}, and \onlinecite{PRL98,PRE99,PRE02}). 

We have found a qualitative and quantitative agreement between our results and the memory functions obtained numerically for spin and energy currents for much larger systems.
The latter analysis allowed us also to explain the origin of the specific scaling of $\lambda_l$. Since this explanation is of perturbative character, we believe
that the validity of the obtained scaling extends down to $\alpha \rightarrow 0$ well beyond the regime which can be inferred directly from bare numerical results.   
     
Within each symmetry sector we have found that the smaller $\lambda_l$ is (roughly understood as a more extended  support of $Q_l$) the larger is the scattering  rate $\gamma_l \alpha^2$. A relevant open question emerges: how many independent scattering rates are introduced by a single perturbation? Since all the scattering rates found in our studies are of the same order of magnitude, this problem may pose a challenge.    

%To illustrate the feasibility of the MCLM and the assumptions on the role of perturbation, we first test the correlations for $j_E$ and $j_s$. 

\acknowledgements
M.M. acknowledges support from the  DEC-2013/09/B/ST3/01659 project of the Polish National Science Center. 
P.P. and T.P. acknowledge support by the program P1-0044 and projects J1-4244 (P.~P.) and J1-5349, N1-0025 (T.~P.) 
of the Slovenian Research Agency.

%%%%%%%%%%%%%%%%%%%%%%%%%%%%%%%%%%%%%%%%

%%%%%%%%%%%%%%%%%%%%%%%%%%%%%%%%%%%%%%%%
%\setcounter{figure}{0}
%\setcounter{equation}{0}
%\renewcommand{\thetable}{S\arabic{table}}
%\renewcommand{\thefigure}{S\arabic{figure}}
%\renewcommand{\theequation}{S\arabic{equation}}
%\renewcommand{\thesection}{S\arabic{section}}
%\label{pagesupp}
%\label{sec:numerics}
%Here we provide more technical details about the choice of the scalar product, the derivation of Eqs. (5-7) from the main text and the details 
%of the numerical calculations of the memory function.\\
\appendix

\section{The choice of  the scalar product}  As an alternative to the scalar product defined in Eq. (\ref{ip1}) one may consider also other scalar products discussed, e.g. in  Ref. \onlinecite{mori65} 
\begin{eqnarray}
(A|B)_1 &=&\frac{1}{2L} \langle A^{\dagger} B +B  A^{\dagger} \rangle \nonumber  \\
&=& \frac{1}{ L} \sum_{mn} A^{*}_{mn} B_{mn} \frac{p(E_n)+p(E_m)}{2},
\label{ip2}
\end{eqnarray} 
or
\begin{eqnarray}
(A|B)_2 &=&\frac{1}{\beta L} \int_{0}^{\beta} {\rm d}x \langle e^{xH}  A^{\dagger} e^{-xH} B\rangle \nonumber  \\
&=& \frac{1}{ L} \sum_{mn} A^{*}_{mn} B_{mn} p(E_n) \frac{e^{\beta(E_n-E_m)}-1}{\beta(E_n-E_m)},
\label{ip3}
\end{eqnarray} 
where $A_{mn}=\langle m | A  | n \rangle $, $H  | n \rangle = E_n  | n \rangle $.
Both these scalar products maintain the essential property, i.e.,  $(\bar{A}^{\tau} |\bar{B}^{\tau} )_1=(\bar{A}^{\tau}|B)_1$ and
$(\bar{A}^{\tau} |\bar{B}^{\tau} )_2=(\bar{A}^{\tau}|B)_2$.  Calculating the scalar products of operators averaged over infinite time--window we find that the only contribution comes from states with equal energies ($E_n=E_m$), hence
\begin{equation}
(\bar{A}|\bar{B})=(\bar{A}|\bar{B})_1=(\bar{A}|\bar{B})_2=\frac{1}{ L} \sum_{E_m=E_n} A^{*}_{mn} B_{mn}  p(E_n).
\end{equation}
Consequently, the stiffness $\frac{1}{L}\langle \bar{A}\bar{A} \rangle$ can be expressed in the same way by all considered scalar products as $(\bar{A}|\bar{A})_{(1,2)}$.
However, when discussing the memory function\cite{mori65} at finite temperature $\beta < \infty$ one should use the scalar product defined in Eq. (\ref{ip3}). 

\section{Support of the approximately conserved quantities}
The orthogonal set of  ACQ consists of operators
\begin{equation}
Q_l  =  \sum_{s} U_{sl} \bar{O}^{\tau}_s,  \label{qdef}
\end{equation}
where the unitary matrix $U$ is defined in Eq. (\ref{eig}) and the norm of $Q_l$ can be found as  
\begin{eqnarray}
||Q_l||^2 & = &\sum_{s,s'} ( U_{sl} \bar{O}^{\tau}_s| U_{s'l} \bar{O}^{\tau}_s)  
%=\sum_{s,s'} U^{*}_{sl} (  \bar{O}^{\tau}_s| \bar{O}^{\tau}_s) U_{s'l'}  \nonumber \\
 =   \sum_{s,s'} U^{\dagger}_{ls} (\bar{O}^{\tau}_s|\bar{O}^{\tau}_{s'}) U_{s'l} = \lambda_{l}.   \nonumber \\
\label{norq}
\end{eqnarray}
We split ACQ into two components $Q_l=Q^M_l+Q^{\perp}_l$, where the former operator is supported on $M$ sites $Q^M_l=\sum_{s'}  (O_{s'}|Q_l) O_{s'} \in {\cal B}^M_L$ while the latter one $(Q^{\perp}_l|O_s)=0$.  Using Eq. (\ref{qdef}), Eq. (\ref{eig}) and the identity $(\bar{A}|\bar{B})=(\bar{A}|\bar{B})$ we find 
\begin{eqnarray}
Q^{M}_l&=& \sum_{s,s'}   (O_{s'}| \bar{O}^{\tau}_s)  U_{sl}  O_{s'} = \sum_{s,s'}   (\bar{O}^{\tau}_{s'}| \bar{O}^{\tau}_s) U_{sl}  O_{s'} \nonumber \\
&&=  \sum_{s'} \lambda_l  U_{s'l}  O_{s'}. 
\end{eqnarray} 
The latter result together with the assumption concerning  the orthonormal basis, $ (O_{s'}|O_s)=\delta_{ss'}$ yields
 \begin{eqnarray}
||Q^M_l||^2&=&( \sum_{s'} \lambda_l  U_{s'l}  O_{s'}| \sum_{s} \lambda_l  U_{sl}  O_{s})  \nonumber \\
& =&\lambda^2_l  \sum_{s,s'}   U^{\dagger}_{ls'}  (O_{s'}|O_s) U_{sl}  =\lambda^2_l \label{norqm}
\end{eqnarray}
Eqs. (\ref{norq}) and (\ref{norqm}) lead to Eq. (\ref{sup}) which relates eigenvalue $\lambda_l$ with the support of $Q_l$.  

\section{Correlation functions and saturated Mazur bound} We consider an operator supported on $M$ sites, $A \in {\cal B}^M_L$ which can be expressed in
terms of basis operators $A =\sum_s a_s O_s$. Using Eq. (\ref{qdef}) we find
\begin{equation}
\bar{A}^{\tau}   = \sum_s a_s \bar{O}^{\tau}_s= \sum_{ls}  a_s   U^{\dagger}_{ls} Q_l \equiv \sum_{l} v_{l} Q_l,  \label{comp1}
\end{equation} 
and
\begin{equation}
 (Q_l|A) = (Q_l|\bar{A}^{\tau}) = v_l \lambda_l.  \label{comp1}
\end{equation} 
Repeating the same calculations for some other operator $B \in {\cal B}^M_L$ we arrive at Eq. (\ref{compg}):
\begin{eqnarray}
(\bar{A}^{\tau}|\bar{B}^{\tau})&=&\sum_{l:\lambda_l \ne 0} \frac{(A|Q_l)}{\lambda_l} (Q_l|Q_l) \frac{(Q_l|B)}{\lambda_l} \nonumber \\
&=&  \sum_{l:\lambda_l \ne 0} \frac{(A|Q_l)(Q_l|B)}{(Q_l|Q_l)} 
\end{eqnarray}

 \bibliography{bib_nongge}

%merlin.mbs apsrev4-1.bst 2010-07-25 4.21a (PWD, AO, DPC) hacked
%Control: key (0)
%Control: author (0) dotless jnrlst
%Control: editor formatted (1) identically to author
%Control: production of article title (0) allowed
%Control: page (1) range
%Control: year (0) verbatim
%Control: production of eprint (0) enabled
\begin{thebibliography}{59}%
\makeatletter
\providecommand \@ifxundefined [1]{%
 \@ifx{#1\undefined}
}%
\providecommand \@ifnum [1]{%
 \ifnum #1\expandafter \@firstoftwo
 \else \expandafter \@secondoftwo
 \fi
}%
\providecommand \@ifx [1]{%
 \ifx #1\expandafter \@firstoftwo
 \else \expandafter \@secondoftwo
 \fi
}%
\providecommand \natexlab [1]{#1}%
\providecommand \enquote  [1]{``#1''}%
\providecommand \bibnamefont  [1]{#1}%
\providecommand \bibfnamefont [1]{#1}%
\providecommand \citenamefont [1]{#1}%
\providecommand \href@noop [0]{\@secondoftwo}%
\providecommand \href [0]{\begingroup \@sanitize@url \@href}%
\providecommand \@href[1]{\@@startlink{#1}\@@href}%
\providecommand \@@href[1]{\endgroup#1\@@endlink}%
\providecommand \@sanitize@url [0]{\catcode `\\12\catcode `\$12\catcode
  `\&12\catcode `\#12\catcode `\^12\catcode `\_12\catcode `\%12\relax}%
\providecommand \@@startlink[1]{}%
\providecommand \@@endlink[0]{}%
\providecommand \url  [0]{\begingroup\@sanitize@url \@url }%
\providecommand \@url [1]{\endgroup\@href {#1}{\urlprefix }}%
\providecommand \urlprefix  [0]{URL }%
\providecommand \Eprint [0]{\href }%
\providecommand \doibase [0]{http://dx.doi.org/}%
\providecommand \selectlanguage [0]{\@gobble}%
\providecommand \bibinfo  [0]{\@secondoftwo}%
\providecommand \bibfield  [0]{\@secondoftwo}%
\providecommand \translation [1]{[#1]}%
\providecommand \BibitemOpen [0]{}%
\providecommand \bibitemStop [0]{}%
\providecommand \bibitemNoStop [0]{.\EOS\space}%
\providecommand \EOS [0]{\spacefactor3000\relax}%
\providecommand \BibitemShut  [1]{\csname bibitem#1\endcsname}%
\let\auto@bib@innerbib\@empty
%</preamble>
\bibitem [{\citenamefont {Polkovnikov}\ \emph {et~al.}(2011)\citenamefont
  {Polkovnikov}, \citenamefont {Sengupta}, \citenamefont {Silva},\ and\
  \citenamefont {Vengalattore}}]{polkrev}%
  \BibitemOpen
  \bibfield  {author} {\bibinfo {author} {\bibfnamefont {Anatoli}\ \bibnamefont
  {Polkovnikov}}, \bibinfo {author} {\bibfnamefont {Krishnendu}\ \bibnamefont
  {Sengupta}}, \bibinfo {author} {\bibfnamefont {Alessandro}\ \bibnamefont
  {Silva}}, \ and\ \bibinfo {author} {\bibfnamefont {Mukund}\ \bibnamefont
  {Vengalattore}},\ }\bibfield  {title} {\enquote {\bibinfo {title}
  {\textit{Colloquium} : Nonequilibrium dynamics of closed interacting quantum
  systems},}\ }\href {\doibase 10.1103/RevModPhys.83.863} {\bibfield  {journal}
  {\bibinfo  {journal} {Rev. Mod. Phys.}\ }\textbf {\bibinfo {volume} {83}},\
  \bibinfo {pages} {863--883} (\bibinfo {year} {2011})}\BibitemShut {NoStop}%
\bibitem [{\citenamefont {Goldstein}\ \emph {et~al.}(2006)\citenamefont
  {Goldstein}, \citenamefont {Lebowitz}, \citenamefont {Tumulka},\ and\
  \citenamefont {Zangh\`\i}}]{Goldstein2006}%
  \BibitemOpen
  \bibfield  {author} {\bibinfo {author} {\bibfnamefont {Sheldon}\ \bibnamefont
  {Goldstein}}, \bibinfo {author} {\bibfnamefont {Joel~L.}\ \bibnamefont
  {Lebowitz}}, \bibinfo {author} {\bibfnamefont {Roderich}\ \bibnamefont
  {Tumulka}}, \ and\ \bibinfo {author} {\bibfnamefont {Nino}\ \bibnamefont
  {Zangh\`\i}},\ }\bibfield  {title} {\enquote {\bibinfo {title} {Canonical
  typicality},}\ }\href {\doibase 10.1103/PhysRevLett.96.050403} {\bibfield
  {journal} {\bibinfo  {journal} {Phys. Rev. Lett.}\ }\textbf {\bibinfo
  {volume} {96}},\ \bibinfo {pages} {050403} (\bibinfo {year}
  {2006})}\BibitemShut {NoStop}%
\bibitem [{\citenamefont {Linden}\ \emph {et~al.}(2009)\citenamefont {Linden},
  \citenamefont {Popescu}, \citenamefont {Short},\ and\ \citenamefont
  {Winter}}]{Linden2009}%
  \BibitemOpen
  \bibfield  {author} {\bibinfo {author} {\bibfnamefont {Noah}\ \bibnamefont
  {Linden}}, \bibinfo {author} {\bibfnamefont {Sandu}\ \bibnamefont {Popescu}},
  \bibinfo {author} {\bibfnamefont {Anthony~J.}\ \bibnamefont {Short}}, \ and\
  \bibinfo {author} {\bibfnamefont {Andreas}\ \bibnamefont {Winter}},\
  }\bibfield  {title} {\enquote {\bibinfo {title} {Quantum mechanical evolution
  towards thermal equilibrium},}\ }\href {\doibase 10.1103/PhysRevE.79.061103}
  {\bibfield  {journal} {\bibinfo  {journal} {Phys. Rev. E}\ }\textbf {\bibinfo
  {volume} {79}},\ \bibinfo {pages} {061103} (\bibinfo {year}
  {2009})}\BibitemShut {NoStop}%
\bibitem [{\citenamefont {Riera}\ \emph {et~al.}(2012)\citenamefont {Riera},
  \citenamefont {Gogolin},\ and\ \citenamefont {Eisert}}]{Riera2012}%
  \BibitemOpen
  \bibfield  {author} {\bibinfo {author} {\bibfnamefont {Arnau}\ \bibnamefont
  {Riera}}, \bibinfo {author} {\bibfnamefont {Christian}\ \bibnamefont
  {Gogolin}}, \ and\ \bibinfo {author} {\bibfnamefont {Jens}\ \bibnamefont
  {Eisert}},\ }\bibfield  {title} {\enquote {\bibinfo {title} {Thermalization
  in nature and on a quantum computer},}\ }\href {\doibase
  10.1103/PhysRevLett.108.080402} {\bibfield  {journal} {\bibinfo  {journal}
  {Phys. Rev. Lett.}\ }\textbf {\bibinfo {volume} {108}},\ \bibinfo {pages}
  {080402} (\bibinfo {year} {2012})}\BibitemShut {NoStop}%
\bibitem [{\citenamefont {Sirker}\ \emph {et~al.}(2014)\citenamefont {Sirker},
  \citenamefont {Konstantinidis}, \citenamefont {Andraschko},\ and\
  \citenamefont {Sedlmayr}}]{sirker2014}%
  \BibitemOpen
  \bibfield  {author} {\bibinfo {author} {\bibfnamefont {J.}~\bibnamefont
  {Sirker}}, \bibinfo {author} {\bibfnamefont {N.~P.}\ \bibnamefont
  {Konstantinidis}}, \bibinfo {author} {\bibfnamefont {F.}~\bibnamefont
  {Andraschko}}, \ and\ \bibinfo {author} {\bibfnamefont {N.}~\bibnamefont
  {Sedlmayr}},\ }\bibfield  {title} {\enquote {\bibinfo {title} {Locality and
  thermalization in closed quantum systems},}\ }\href {\doibase
  10.1103/PhysRevA.89.042104} {\bibfield  {journal} {\bibinfo  {journal} {Phys.
  Rev. A}\ }\textbf {\bibinfo {volume} {89}},\ \bibinfo {pages} {042104}
  (\bibinfo {year} {2014})}\BibitemShut {NoStop}%
\bibitem [{\citenamefont {Konstantinidis}(2015)}]{konstantinidis2015}%
  \BibitemOpen
  \bibfield  {author} {\bibinfo {author} {\bibfnamefont {N.~P.}\ \bibnamefont
  {Konstantinidis}},\ }\bibfield  {title} {\enquote {\bibinfo {title}
  {Thermalization away from integrability and the role of operator off-diagonal
  elements},}\ }\href {\doibase 10.1103/PhysRevE.91.052111} {\bibfield
  {journal} {\bibinfo  {journal} {Phys. Rev. E}\ }\textbf {\bibinfo {volume}
  {91}},\ \bibinfo {pages} {052111} (\bibinfo {year} {2015})}\BibitemShut
  {NoStop}%
\bibitem [{\citenamefont {Hershfield}(1993)}]{Hershfield1993}%
  \BibitemOpen
  \bibfield  {author} {\bibinfo {author} {\bibfnamefont {Selman}\ \bibnamefont
  {Hershfield}},\ }\bibfield  {title} {\enquote {\bibinfo {title}
  {Reformulation of steady state nonequilibrium quantum statistical
  mechanics},}\ }\href {\doibase 10.1103/PhysRevLett.70.2134} {\bibfield
  {journal} {\bibinfo  {journal} {Phys. Rev. Lett.}\ }\textbf {\bibinfo
  {volume} {70}},\ \bibinfo {pages} {2134--2137} (\bibinfo {year}
  {1993})}\BibitemShut {NoStop}%
\bibitem [{\citenamefont {Doyon}\ and\ \citenamefont
  {Andrei}(2006)}]{Natan2006}%
  \BibitemOpen
  \bibfield  {author} {\bibinfo {author} {\bibfnamefont {Benjamin}\
  \bibnamefont {Doyon}}\ and\ \bibinfo {author} {\bibfnamefont {Natan}\
  \bibnamefont {Andrei}},\ }\bibfield  {title} {\enquote {\bibinfo {title}
  {Universal aspects of nonequilibrium currents in a quantum dot},}\ }\href
  {\doibase 10.1103/PhysRevB.73.245326} {\bibfield  {journal} {\bibinfo
  {journal} {Phys. Rev. B}\ }\textbf {\bibinfo {volume} {73}},\ \bibinfo
  {pages} {245326} (\bibinfo {year} {2006})}\BibitemShut {NoStop}%
\bibitem [{\citenamefont {Santos}\ \emph {et~al.}(2011)\citenamefont {Santos},
  \citenamefont {Polkovnikov},\ and\ \citenamefont {Rigol}}]{santos2011}%
  \BibitemOpen
  \bibfield  {author} {\bibinfo {author} {\bibfnamefont {Lea~F.}\ \bibnamefont
  {Santos}}, \bibinfo {author} {\bibfnamefont {Anatoli}\ \bibnamefont
  {Polkovnikov}}, \ and\ \bibinfo {author} {\bibfnamefont {Marcos}\
  \bibnamefont {Rigol}},\ }\bibfield  {title} {\enquote {\bibinfo {title}
  {Entropy of isolated quantum systems after a quench},}\ }\href {\doibase
  10.1103/PhysRevLett.107.040601} {\bibfield  {journal} {\bibinfo  {journal}
  {Phys. Rev. Lett.}\ }\textbf {\bibinfo {volume} {107}},\ \bibinfo {pages}
  {040601} (\bibinfo {year} {2011})}\BibitemShut {NoStop}%
\bibitem [{\citenamefont {Mierzejewski}\ \emph {et~al.}(2013)\citenamefont
  {Mierzejewski}, \citenamefont {Prosen}, \citenamefont {Crivelli},\ and\
  \citenamefont {Prelov\ifmmode~\check{s}\else \v{s}\fi{}ek}}]{our2013}%
  \BibitemOpen
  \bibfield  {author} {\bibinfo {author} {\bibfnamefont {M.}~\bibnamefont
  {Mierzejewski}}, \bibinfo {author} {\bibfnamefont {T.}~\bibnamefont
  {Prosen}}, \bibinfo {author} {\bibfnamefont {D.}~\bibnamefont {Crivelli}}, \
  and\ \bibinfo {author} {\bibfnamefont {P.}~\bibnamefont
  {Prelov\ifmmode~\check{s}\else \v{s}\fi{}ek}},\ }\bibfield  {title} {\enquote
  {\bibinfo {title} {Eigenvalue statistics of reduced density matrix during
  driving and relaxation},}\ }\href {\doibase 10.1103/PhysRevLett.110.200602}
  {\bibfield  {journal} {\bibinfo  {journal} {Phys. Rev. Lett.}\ }\textbf
  {\bibinfo {volume} {110}},\ \bibinfo {pages} {200602} (\bibinfo {year}
  {2013})}\BibitemShut {NoStop}%
\bibitem [{\citenamefont {Rigol}\ \emph {et~al.}(2007)\citenamefont {Rigol},
  \citenamefont {Dunjko}, \citenamefont {Yurovsky},\ and\ \citenamefont
  {Olshanii}}]{gge}%
  \BibitemOpen
  \bibfield  {author} {\bibinfo {author} {\bibfnamefont {Marcos}\ \bibnamefont
  {Rigol}}, \bibinfo {author} {\bibfnamefont {Vanja}\ \bibnamefont {Dunjko}},
  \bibinfo {author} {\bibfnamefont {Vladimir}\ \bibnamefont {Yurovsky}}, \ and\
  \bibinfo {author} {\bibfnamefont {Maxim}\ \bibnamefont {Olshanii}},\
  }\bibfield  {title} {\enquote {\bibinfo {title} {Relaxation in a completely
  integrable many-body quantum system: An \textit{Ab~Initio} study of the
  dynamics of the highly excited states of $\mathrm{1D}$ lattice hard-core
  bosons},}\ }\href {\doibase 10.1103/PhysRevLett.98.050405} {\bibfield
  {journal} {\bibinfo  {journal} {Phys. Rev. Lett.}\ }\textbf {\bibinfo
  {volume} {98}},\ \bibinfo {pages} {050405} (\bibinfo {year}
  {2007})}\BibitemShut {NoStop}%
\bibitem [{\citenamefont {Kollar}\ \emph {et~al.}(2011)\citenamefont {Kollar},
  \citenamefont {Wolf},\ and\ \citenamefont {Eckstein}}]{Eckstein2012}%
  \BibitemOpen
  \bibfield  {author} {\bibinfo {author} {\bibfnamefont {Marcus}\ \bibnamefont
  {Kollar}}, \bibinfo {author} {\bibfnamefont {F.~Alexander}\ \bibnamefont
  {Wolf}}, \ and\ \bibinfo {author} {\bibfnamefont {Martin}\ \bibnamefont
  {Eckstein}},\ }\bibfield  {title} {\enquote {\bibinfo {title} {Generalized
  $\mathrm{Gibbs}$ ensemble prediction of prethermalization plateaus and their
  relation to nonthermal steady states in integrable systems},}\ }\href
  {\doibase 10.1103/PhysRevB.84.054304} {\bibfield  {journal} {\bibinfo
  {journal} {Phys. Rev. B}\ }\textbf {\bibinfo {volume} {84}},\ \bibinfo
  {pages} {054304} (\bibinfo {year} {2011})}\BibitemShut {NoStop}%
\bibitem [{\citenamefont {Cassidy}\ \emph {et~al.}(2011)\citenamefont
  {Cassidy}, \citenamefont {Clark},\ and\ \citenamefont {Rigol}}]{Cassidy2011}%
  \BibitemOpen
  \bibfield  {author} {\bibinfo {author} {\bibfnamefont {Amy~C.}\ \bibnamefont
  {Cassidy}}, \bibinfo {author} {\bibfnamefont {Charles~W.}\ \bibnamefont
  {Clark}}, \ and\ \bibinfo {author} {\bibfnamefont {Marcos}\ \bibnamefont
  {Rigol}},\ }\bibfield  {title} {\enquote {\bibinfo {title} {Generalized
  thermalization in an integrable lattice system},}\ }\href {\doibase
  10.1103/PhysRevLett.106.140405} {\bibfield  {journal} {\bibinfo  {journal}
  {Phys. Rev. Lett.}\ }\textbf {\bibinfo {volume} {106}},\ \bibinfo {pages}
  {140405} (\bibinfo {year} {2011})}\BibitemShut {NoStop}%
\bibitem [{\citenamefont {Gogolin}\ \emph {et~al.}(2011)\citenamefont
  {Gogolin}, \citenamefont {M\"uller},\ and\ \citenamefont
  {Eisert}}]{Gogolin2011}%
  \BibitemOpen
  \bibfield  {author} {\bibinfo {author} {\bibfnamefont {Christian}\
  \bibnamefont {Gogolin}}, \bibinfo {author} {\bibfnamefont {Markus~P.}\
  \bibnamefont {M\"uller}}, \ and\ \bibinfo {author} {\bibfnamefont {Jens}\
  \bibnamefont {Eisert}},\ }\bibfield  {title} {\enquote {\bibinfo {title}
  {Absence of thermalization in nonintegrable systems},}\ }\href {\doibase
  10.1103/PhysRevLett.106.040401} {\bibfield  {journal} {\bibinfo  {journal}
  {Phys. Rev. Lett.}\ }\textbf {\bibinfo {volume} {106}},\ \bibinfo {pages}
  {040401} (\bibinfo {year} {2011})}\BibitemShut {NoStop}%
\bibitem [{\citenamefont {Fagotti}\ \emph {et~al.}(2014)\citenamefont
  {Fagotti}, \citenamefont {Collura}, \citenamefont {Essler},\ and\
  \citenamefont {Calabrese}}]{essler2014}%
  \BibitemOpen
  \bibfield  {author} {\bibinfo {author} {\bibfnamefont {Maurizio}\
  \bibnamefont {Fagotti}}, \bibinfo {author} {\bibfnamefont {Mario}\
  \bibnamefont {Collura}}, \bibinfo {author} {\bibfnamefont {Fabian H.~L.}\
  \bibnamefont {Essler}}, \ and\ \bibinfo {author} {\bibfnamefont {Pasquale}\
  \bibnamefont {Calabrese}},\ }\bibfield  {title} {\enquote {\bibinfo {title}
  {Relaxation after quantum quenches in the spin-$\frac{1}{2}$
  $\mathrm{Heisenberg}$ $\mathrm{XXZ}$ chain},}\ }\href {\doibase
  10.1103/PhysRevB.89.125101} {\bibfield  {journal} {\bibinfo  {journal} {Phys.
  Rev. B}\ }\textbf {\bibinfo {volume} {89}},\ \bibinfo {pages} {125101}
  (\bibinfo {year} {2014})}\BibitemShut {NoStop}%
\bibitem [{\citenamefont {Rigol}(2014)}]{rigol2014}%
  \BibitemOpen
  \bibfield  {author} {\bibinfo {author} {\bibfnamefont {Marcos}\ \bibnamefont
  {Rigol}},\ }\bibfield  {title} {\enquote {\bibinfo {title} {Quantum quenches
  in the thermodynamic limit. $\mathrm{II}$. initial ground states},}\ }\href
  {\doibase 10.1103/PhysRevE.90.031301} {\bibfield  {journal} {\bibinfo
  {journal} {Phys. Rev. E}\ }\textbf {\bibinfo {volume} {90}},\ \bibinfo
  {pages} {031301} (\bibinfo {year} {2014})}\BibitemShut {NoStop}%
\bibitem [{\citenamefont {Mierzejewski}\ \emph {et~al.}(2014)\citenamefont
  {Mierzejewski}, \citenamefont {Prelov\ifmmode~\check{s}\else \v{s}\fi{}ek},\
  and\ \citenamefont {Prosen}}]{nongge1}%
  \BibitemOpen
  \bibfield  {author} {\bibinfo {author} {\bibfnamefont {Marcin}\ \bibnamefont
  {Mierzejewski}}, \bibinfo {author} {\bibfnamefont {Peter}\ \bibnamefont
  {Prelov\ifmmode~\check{s}\else \v{s}\fi{}ek}}, \ and\ \bibinfo {author}
  {\bibfnamefont {Toma\ifmmode\check{z}\else\v{z}\fi{}}\ \bibnamefont
  {Prosen}},\ }\bibfield  {title} {\enquote {\bibinfo {title} {Breakdown of the
  generalized $\mathrm{Gibbs}$ ensemble for current-generating quenches},}\
  }\href {\doibase 10.1103/PhysRevLett.113.020602} {\bibfield  {journal}
  {\bibinfo  {journal} {Phys. Rev. Lett.}\ }\textbf {\bibinfo {volume} {113}},\
  \bibinfo {pages} {020602} (\bibinfo {year} {2014})}\BibitemShut {NoStop}%
\bibitem [{\citenamefont {Pozsgay}\ \emph {et~al.}(2014)\citenamefont
  {Pozsgay}, \citenamefont {Mesty\'an}, \citenamefont {Werner}, \citenamefont
  {Kormos}, \citenamefont {Zar\'and},\ and\ \citenamefont
  {Tak\'acs}}]{nongge2}%
  \BibitemOpen
  \bibfield  {author} {\bibinfo {author} {\bibfnamefont {B.}~\bibnamefont
  {Pozsgay}}, \bibinfo {author} {\bibfnamefont {M.}~\bibnamefont {Mesty\'an}},
  \bibinfo {author} {\bibfnamefont {M.~A.}\ \bibnamefont {Werner}}, \bibinfo
  {author} {\bibfnamefont {M.}~\bibnamefont {Kormos}}, \bibinfo {author}
  {\bibfnamefont {G.}~\bibnamefont {Zar\'and}}, \ and\ \bibinfo {author}
  {\bibfnamefont {G.}~\bibnamefont {Tak\'acs}},\ }\bibfield  {title} {\enquote
  {\bibinfo {title} {Correlations after quantum quenches in the $\mathrm{XXZ}$
  spin chain: Failure of the generalized $\mathrm{Gibbs}$ ensemble},}\ }\href
  {\doibase 10.1103/PhysRevLett.113.117203} {\bibfield  {journal} {\bibinfo
  {journal} {Phys. Rev. Lett.}\ }\textbf {\bibinfo {volume} {113}},\ \bibinfo
  {pages} {117203} (\bibinfo {year} {2014})}\BibitemShut {NoStop}%
\bibitem [{\citenamefont {Goldstein}\ and\ \citenamefont
  {Andrei}(2014)}]{nongge3}%
  \BibitemOpen
  \bibfield  {author} {\bibinfo {author} {\bibfnamefont {Garry}\ \bibnamefont
  {Goldstein}}\ and\ \bibinfo {author} {\bibfnamefont {Natan}\ \bibnamefont
  {Andrei}},\ }\bibfield  {title} {\enquote {\bibinfo {title} {Failure of the
  local generalized $\mathrm{Gibbs}$ ensemble for integrable models with bound
  states},}\ }\href {\doibase 10.1103/PhysRevA.90.043625} {\bibfield  {journal}
  {\bibinfo  {journal} {Phys. Rev. A}\ }\textbf {\bibinfo {volume} {90}},\
  \bibinfo {pages} {043625} (\bibinfo {year} {2014})}\BibitemShut {NoStop}%
\bibitem [{\citenamefont {Prosen}(2011{\natexlab{a}})}]{tomaz_quasilocal11}%
  \BibitemOpen
  \bibfield  {author} {\bibinfo {author} {\bibfnamefont {Toma\v{z}}\
  \bibnamefont {Prosen}},\ }\bibfield  {title} {\enquote {\bibinfo {title}
  {Open $\mathrm{XXZ}$ spin chain: Nonequilibrium steady state and a strict
  bound on ballistic transport},}\ }\href {\doibase
  10.1103/PhysRevLett.106.217206} {\bibfield  {journal} {\bibinfo  {journal}
  {Phys. Rev. Lett.}\ }\textbf {\bibinfo {volume} {106}},\ \bibinfo {pages}
  {217206} (\bibinfo {year} {2011}{\natexlab{a}})}\BibitemShut {NoStop}%
\bibitem [{\citenamefont {Prosen}\ and\ \citenamefont
  {Ilievski}(2013)}]{tomaz_quasilocal13}%
  \BibitemOpen
  \bibfield  {author} {\bibinfo {author} {\bibfnamefont {Toma\v{z}}\
  \bibnamefont {Prosen}}\ and\ \bibinfo {author} {\bibfnamefont {Enej}\
  \bibnamefont {Ilievski}},\ }\bibfield  {title} {\enquote {\bibinfo {title}
  {Families of quasilocal conservation laws and quantum spin transport},}\
  }\href {\doibase 10.1103/PhysRevLett.111.057203} {\bibfield  {journal}
  {\bibinfo  {journal} {Phys. Rev. Lett.}\ }\textbf {\bibinfo {volume} {111}},\
  \bibinfo {pages} {057203} (\bibinfo {year} {2013})}\BibitemShut {NoStop}%
\bibitem [{\citenamefont {Prosen}(2014)}]{tomaz_quasilocal14}%
  \BibitemOpen
  \bibfield  {author} {\bibinfo {author} {\bibfnamefont {Tomaz}\ \bibnamefont
  {Prosen}},\ }\bibfield  {title} {\enquote {\bibinfo {title} {Quasilocal
  conservation laws in $\mathrm{XXZ}$ spin-1/2 chains: Open, periodic and
  twisted boundary conditions},}\ }\href {\doibase
  http://dx.doi.org/10.1016/j.nuclphysb.2014.07.024} {\bibfield  {journal}
  {\bibinfo  {journal} {Nuclear Physics B}\ }\textbf {\bibinfo {volume}
  {886}},\ \bibinfo {pages} {1177 -- 1198} (\bibinfo {year}
  {2014})}\BibitemShut {NoStop}%
\bibitem [{\citenamefont {Pereira}\ \emph {et~al.}(2014)\citenamefont
  {Pereira}, \citenamefont {Pasquier}, \citenamefont {Sirker},\ and\
  \citenamefont {Affleck}}]{affleck14}%
  \BibitemOpen
  \bibfield  {author} {\bibinfo {author} {\bibfnamefont {R.~G.}\ \bibnamefont
  {Pereira}}, \bibinfo {author} {\bibfnamefont {V.}~\bibnamefont {Pasquier}},
  \bibinfo {author} {\bibfnamefont {J.}~\bibnamefont {Sirker}}, \ and\ \bibinfo
  {author} {\bibfnamefont {I.}~\bibnamefont {Affleck}},\ }\bibfield  {title}
  {\enquote {\bibinfo {title} {Exactly conserved quasilocal operators for the
  $\mathrm{XXZ}$ spin chain},}\ }\href
  {http://stacks.iop.org/1742-5468/2014/i=9/a=P09037} {\bibfield  {journal}
  {\bibinfo  {journal} {Journal of Statistical Mechanics: Theory and
  Experiment}\ }\textbf {\bibinfo {volume} {2014}},\ \bibinfo {pages} {P09037}
  (\bibinfo {year} {2014})}\BibitemShut {NoStop}%
\bibitem [{\citenamefont {Mierzejewski}\ \emph {et~al.}(2015)\citenamefont
  {Mierzejewski}, \citenamefont {Prelov\ifmmode~\check{s}\else \v{s}\fi{}ek},\
  and\ \citenamefont {Prosen}}]{mierzejewski2015}%
  \BibitemOpen
  \bibfield  {author} {\bibinfo {author} {\bibfnamefont {Marcin}\ \bibnamefont
  {Mierzejewski}}, \bibinfo {author} {\bibfnamefont {Peter}\ \bibnamefont
  {Prelov\ifmmode~\check{s}\else \v{s}\fi{}ek}}, \ and\ \bibinfo {author}
  {\bibfnamefont {Toma\v{z}}\ \bibnamefont {Prosen}},\ }\bibfield  {title}
  {\enquote {\bibinfo {title} {Identifying local and quasilocal conserved
  quantities in integrable systems},}\ }\href {\doibase
  10.1103/PhysRevLett.114.140601} {\bibfield  {journal} {\bibinfo  {journal}
  {Phys. Rev. Lett.}\ }\textbf {\bibinfo {volume} {114}},\ \bibinfo {pages}
  {140601} (\bibinfo {year} {2015})}\BibitemShut {NoStop}%
\bibitem [{\citenamefont {Ilievski}\ \emph {et~al.}(2015)\citenamefont
  {Ilievski}, \citenamefont {De~Nardis}, \citenamefont {Wouters}, \citenamefont
  {Caux}, \citenamefont {Essler},\ and\ \citenamefont {Prosen}}]{ilievski}%
  \BibitemOpen
  \bibfield  {author} {\bibinfo {author} {\bibfnamefont {E.}~\bibnamefont
  {Ilievski}}, \bibinfo {author} {\bibfnamefont {J.}~\bibnamefont {De~Nardis}},
  \bibinfo {author} {\bibfnamefont {B.}~\bibnamefont {Wouters}}, \bibinfo
  {author} {\bibfnamefont {J.-S.}\ \bibnamefont {Caux}}, \bibinfo {author}
  {\bibfnamefont {F.~H.~L.}\ \bibnamefont {Essler}}, \ and\ \bibinfo {author}
  {\bibfnamefont {T.}~\bibnamefont {Prosen}},\ }\bibfield  {title} {\enquote
  {\bibinfo {title} {Complete generalized gibbs ensembles in interacting
  theories},}\ }\href {http://arxiv.org/abs/1507.02993} {\bibfield  {journal}
  {\bibinfo  {journal} {arXiv:1507.02993}\ } (\bibinfo {year}
  {2015})}\BibitemShut {NoStop}%
\bibitem [{\citenamefont {Tetelman}(1982)}]{tetelman}%
  \BibitemOpen
  \bibfield  {author} {\bibinfo {author} {\bibfnamefont {M.G.}\ \bibnamefont
  {Tetelman}},\ }\bibfield  {title} {\enquote {\bibinfo {title} {Lorentz group
  for two-dimensional integrable lattice systems},}\ }\href@noop {} {\bibfield
  {journal} {\bibinfo  {journal} {Sov. Phys. JETP}\ }\textbf {\bibinfo {volume}
  {55}},\ \bibinfo {pages} {306} (\bibinfo {year} {1982})}\BibitemShut
  {NoStop}%
\bibitem [{\citenamefont {Grabowski}\ and\ \citenamefont
  {Mathieu}(1995)}]{grabowski}%
  \BibitemOpen
  \bibfield  {author} {\bibinfo {author} {\bibfnamefont {M.~P.}\ \bibnamefont
  {Grabowski}}\ and\ \bibinfo {author} {\bibfnamefont {P.}~\bibnamefont
  {Mathieu}},\ }\bibfield  {title} {\enquote {\bibinfo {title} {Structure of
  the conservation laws in quantum integrable spin chains with short range
  interactions},}\ }\href@noop {} {\bibfield  {journal} {\bibinfo  {journal}
  {Ann. Phys. (N.Y.)}\ }\textbf {\bibinfo {volume} {243}},\ \bibinfo {pages}
  {299} (\bibinfo {year} {1995})}\BibitemShut {NoStop}%
\bibitem [{\citenamefont {Zotos}\ and\ \citenamefont
  {Prelov\ifmmode~\check{s}\else \v{s}\fi{}ek}(1996)}]{zotos1996}%
  \BibitemOpen
  \bibfield  {author} {\bibinfo {author} {\bibfnamefont {X.}~\bibnamefont
  {Zotos}}\ and\ \bibinfo {author} {\bibfnamefont {P.}~\bibnamefont
  {Prelov\ifmmode~\check{s}\else \v{s}\fi{}ek}},\ }\bibfield  {title} {\enquote
  {\bibinfo {title} {Evidence for ideal insulating or conducting state in a
  one-dimensional integrable system},}\ }\href {\doibase
  10.1103/PhysRevB.53.983} {\bibfield  {journal} {\bibinfo  {journal} {Phys.
  Rev. B}\ }\textbf {\bibinfo {volume} {53}},\ \bibinfo {pages} {983--986}
  (\bibinfo {year} {1996})}\BibitemShut {NoStop}%
\bibitem [{\citenamefont {Zotos}\ \emph {et~al.}(1997)\citenamefont {Zotos},
  \citenamefont {Naef},\ and\ \citenamefont {Prelovsek}}]{zotos1997}%
  \BibitemOpen
  \bibfield  {author} {\bibinfo {author} {\bibfnamefont {X.}~\bibnamefont
  {Zotos}}, \bibinfo {author} {\bibfnamefont {F.}~\bibnamefont {Naef}}, \ and\
  \bibinfo {author} {\bibfnamefont {P.}~\bibnamefont {Prelovsek}},\ }\bibfield
  {title} {\enquote {\bibinfo {title} {Transport and conservation laws},}\
  }\href {\doibase 10.1103/PhysRevB.55.11029} {\bibfield  {journal} {\bibinfo
  {journal} {Phys. Rev. B}\ }\textbf {\bibinfo {volume} {55}},\ \bibinfo
  {pages} {11029--11032} (\bibinfo {year} {1997})}\BibitemShut {NoStop}%
\bibitem [{\citenamefont {Zotos}(1999)}]{zotos1999}%
  \BibitemOpen
  \bibfield  {author} {\bibinfo {author} {\bibfnamefont {X.}~\bibnamefont
  {Zotos}},\ }\bibfield  {title} {\enquote {\bibinfo {title} {Finite
  temperature $\mathrm{Drude}$ weight of the one-dimensional spin- $1/2$
  $\mathrm{Heisenberg}$ model},}\ }\href {\doibase 10.1103/PhysRevLett.82.1764}
  {\bibfield  {journal} {\bibinfo  {journal} {Phys. Rev. Lett.}\ }\textbf
  {\bibinfo {volume} {82}},\ \bibinfo {pages} {1764--1767} (\bibinfo {year}
  {1999})}\BibitemShut {NoStop}%
\bibitem [{\citenamefont {Hawkins}\ \emph {et~al.}(2008)\citenamefont
  {Hawkins}, \citenamefont {Long},\ and\ \citenamefont {Zotos}}]{zotos_unpub}%
  \BibitemOpen
  \bibfield  {author} {\bibinfo {author} {\bibfnamefont {M.S.}\ \bibnamefont
  {Hawkins}}, \bibinfo {author} {\bibfnamefont {M.W.}\ \bibnamefont {Long}}, \
  and\ \bibinfo {author} {\bibfnamefont {X.}~\bibnamefont {Zotos}},\ }\bibfield
   {title} {\enquote {\bibinfo {title} {Long-time asymptotics and conservation
  laws in integrable systems},}\ }\href {http://arxiv.org/abs/0812.3096}
  {\bibfield  {journal} {\bibinfo  {journal} {arXiv:0812.3096v1}\ } (\bibinfo
  {year} {2008})}\BibitemShut {NoStop}%
\bibitem [{\citenamefont {Benz}\ \emph {et~al.}(2005)\citenamefont {Benz},
  \citenamefont {Fukui}, \citenamefont {Kl{\"u}mper},\ and\ \citenamefont
  {Scheeren}}]{benz}%
  \BibitemOpen
  \bibfield  {author} {\bibinfo {author} {\bibfnamefont {J.}~\bibnamefont
  {Benz}}, \bibinfo {author} {\bibfnamefont {T.}~\bibnamefont {Fukui}},
  \bibinfo {author} {\bibfnamefont {A.}~\bibnamefont {Kl{\"u}mper}}, \ and\
  \bibinfo {author} {\bibfnamefont {C.}~\bibnamefont {Scheeren}},\ }\bibfield
  {title} {\enquote {\bibinfo {title} {On the finite temperature
  $\mathrm{Drude}$ weight of the anisotropic $\mathrm{Heisenberg}$ chain},}\
  }\href {\doibase 10.1143/JPSJS.74S.181} {\bibfield  {journal} {\bibinfo
  {journal} {Journal of the Physical Society of Japan}\ }\textbf {\bibinfo
  {volume} {74}},\ \bibinfo {pages} {181--190} (\bibinfo {year} {2005})},\
  \Eprint
  {http://arxiv.org/abs/http://journals.jps.jp/doi/pdf/10.1143/JPSJS.74S.181}
  {http://journals.jps.jp/doi/pdf/10.1143/JPSJS.74S.181} \BibitemShut {NoStop}%
\bibitem [{\citenamefont {Heidrich-Meisner}\ \emph {et~al.}(2007)\citenamefont
  {Heidrich-Meisner}, \citenamefont {Honecker},\ and\ \citenamefont
  {Brenig}}]{review2007}%
  \BibitemOpen
  \bibfield  {author} {\bibinfo {author} {\bibfnamefont {F.}~\bibnamefont
  {Heidrich-Meisner}}, \bibinfo {author} {\bibfnamefont {A.}~\bibnamefont
  {Honecker}}, \ and\ \bibinfo {author} {\bibfnamefont {W.}~\bibnamefont
  {Brenig}},\ }\bibfield  {title} {\enquote {\bibinfo {title} {Transport in
  quasi one-dimensional spin-1/2 systems},}\ }\href {\doibase
  10.1140/epjst/e2007-00369-2} {\bibfield  {journal} {\bibinfo  {journal} {The
  European Physical Journal Special Topics}\ }\textbf {\bibinfo {volume}
  {151}},\ \bibinfo {pages} {135--145} (\bibinfo {year} {2007})}\BibitemShut
  {NoStop}%
\bibitem [{\citenamefont {Rigol}\ and\ \citenamefont
  {Shastry}(2008)}]{shastry}%
  \BibitemOpen
  \bibfield  {author} {\bibinfo {author} {\bibfnamefont {Marcos}\ \bibnamefont
  {Rigol}}\ and\ \bibinfo {author} {\bibfnamefont {B.~Sriram}\ \bibnamefont
  {Shastry}},\ }\bibfield  {title} {\enquote {\bibinfo {title} {Drude weight in
  systems with open boundary conditions},}\ }\href {\doibase
  10.1103/PhysRevB.77.161101} {\bibfield  {journal} {\bibinfo  {journal} {Phys.
  Rev. B}\ }\textbf {\bibinfo {volume} {77}},\ \bibinfo {pages} {161101}
  (\bibinfo {year} {2008})}\BibitemShut {NoStop}%
\bibitem [{\citenamefont {Herbrych}\ \emph {et~al.}(2011)\citenamefont
  {Herbrych}, \citenamefont {Prelov\v{s}ek},\ and\ \citenamefont
  {Zotos}}]{herbrych2011}%
  \BibitemOpen
  \bibfield  {author} {\bibinfo {author} {\bibfnamefont {J.}~\bibnamefont
  {Herbrych}}, \bibinfo {author} {\bibfnamefont {P.}~\bibnamefont
  {Prelov\v{s}ek}}, \ and\ \bibinfo {author} {\bibfnamefont {X.}~\bibnamefont
  {Zotos}},\ }\bibfield  {title} {\enquote {\bibinfo {title}
  {Finite-temperature $\mathrm{Drude}$ weight within the anisotropic heisenberg
  chain},}\ }\href {\doibase 10.1103/PhysRevB.84.155125} {\bibfield  {journal}
  {\bibinfo  {journal} {Phys. Rev. B}\ }\textbf {\bibinfo {volume} {84}},\
  \bibinfo {pages} {155125} (\bibinfo {year} {2011})}\BibitemShut {NoStop}%
\bibitem [{\citenamefont {\v{Z}nidari\v{c}}(2011)}]{Marko2011}%
  \BibitemOpen
  \bibfield  {author} {\bibinfo {author} {\bibfnamefont {Marko}\ \bibnamefont
  {\v{Z}nidari\v{c}}},\ }\bibfield  {title} {\enquote {\bibinfo {title} {Spin
  transport in a one-dimensional anisotropic $\mathrm{Heisenberg}$ model},}\
  }\href {\doibase 10.1103/PhysRevLett.106.220601} {\bibfield  {journal}
  {\bibinfo  {journal} {Phys. Rev. Lett.}\ }\textbf {\bibinfo {volume} {106}},\
  \bibinfo {pages} {220601} (\bibinfo {year} {2011})}\BibitemShut {NoStop}%
\bibitem [{\citenamefont {Sirker}\ \emph {et~al.}(2009)\citenamefont {Sirker},
  \citenamefont {Pereira},\ and\ \citenamefont {Affleck}}]{Sirker2009}%
  \BibitemOpen
  \bibfield  {author} {\bibinfo {author} {\bibfnamefont {J.}~\bibnamefont
  {Sirker}}, \bibinfo {author} {\bibfnamefont {R.~G.}\ \bibnamefont {Pereira}},
  \ and\ \bibinfo {author} {\bibfnamefont {I.}~\bibnamefont {Affleck}},\
  }\bibfield  {title} {\enquote {\bibinfo {title} {Diffusion and ballistic
  transport in one-dimensional quantum systems},}\ }\href {\doibase
  10.1103/PhysRevLett.103.216602} {\bibfield  {journal} {\bibinfo  {journal}
  {Phys. Rev. Lett.}\ }\textbf {\bibinfo {volume} {103}},\ \bibinfo {pages}
  {216602} (\bibinfo {year} {2009})}\BibitemShut {NoStop}%
\bibitem [{\citenamefont {Steinigeweg}\ \emph {et~al.}(2014)\citenamefont
  {Steinigeweg}, \citenamefont {Gemmer},\ and\ \citenamefont
  {Brenig}}]{robin2013}%
  \BibitemOpen
  \bibfield  {author} {\bibinfo {author} {\bibfnamefont {Robin}\ \bibnamefont
  {Steinigeweg}}, \bibinfo {author} {\bibfnamefont {Jochen}\ \bibnamefont
  {Gemmer}}, \ and\ \bibinfo {author} {\bibfnamefont {Wolfram}\ \bibnamefont
  {Brenig}},\ }\bibfield  {title} {\enquote {\bibinfo {title} {Spin-current
  autocorrelations from single pure-state propagation},}\ }\href {\doibase
  10.1103/PhysRevLett.112.120601} {\bibfield  {journal} {\bibinfo  {journal}
  {Phys. Rev. Lett.}\ }\textbf {\bibinfo {volume} {112}},\ \bibinfo {pages}
  {120601} (\bibinfo {year} {2014})}\BibitemShut {NoStop}%
\bibitem [{\citenamefont {Prosen}(2011{\natexlab{b}})}]{Tomaz2011}%
  \BibitemOpen
  \bibfield  {author} {\bibinfo {author} {\bibfnamefont {Tomaz}\ \bibnamefont
  {Prosen}},\ }\bibfield  {title} {\enquote {\bibinfo {title} {Exact
  nonequilibrium steady state of a strongly driven open $\mathrm{XXZ}$
  chain},}\ }\href {\doibase 10.1103/PhysRevLett.107.137201} {\bibfield
  {journal} {\bibinfo  {journal} {Phys. Rev. Lett.}\ }\textbf {\bibinfo
  {volume} {107}},\ \bibinfo {pages} {137201} (\bibinfo {year}
  {2011}{\natexlab{b}})}\BibitemShut {NoStop}%
\bibitem [{\citenamefont {Steinigeweg}\ \emph {et~al.}(2012)\citenamefont
  {Steinigeweg}, \citenamefont {Herbrych}, \citenamefont {Prelov\v{s}ek},\ and\
  \citenamefont {Mierzejewski}}]{my3}%
  \BibitemOpen
  \bibfield  {author} {\bibinfo {author} {\bibfnamefont {R.}~\bibnamefont
  {Steinigeweg}}, \bibinfo {author} {\bibfnamefont {J.}~\bibnamefont
  {Herbrych}}, \bibinfo {author} {\bibfnamefont {P.}~\bibnamefont
  {Prelov\v{s}ek}}, \ and\ \bibinfo {author} {\bibfnamefont {M.}~\bibnamefont
  {Mierzejewski}},\ }\bibfield  {title} {\enquote {\bibinfo {title}
  {Coexistence of anomalous and normal diffusion in integrable mott
  insulators},}\ }\href {\doibase 10.1103/PhysRevB.85.214409} {\bibfield
  {journal} {\bibinfo  {journal} {Phys. Rev. B}\ }\textbf {\bibinfo {volume}
  {85}},\ \bibinfo {pages} {214409} (\bibinfo {year} {2012})}\BibitemShut
  {NoStop}%
\bibitem [{\citenamefont {Vidmar}\ \emph {et~al.}(2013)\citenamefont {Vidmar},
  \citenamefont {Langer}, \citenamefont {McCulloch}, \citenamefont {Schneider},
  \citenamefont {Schollw\"ock},\ and\ \citenamefont
  {Heidrich-Meisner}}]{vidmar2013}%
  \BibitemOpen
  \bibfield  {author} {\bibinfo {author} {\bibfnamefont {L.}~\bibnamefont
  {Vidmar}}, \bibinfo {author} {\bibfnamefont {S.}~\bibnamefont {Langer}},
  \bibinfo {author} {\bibfnamefont {I.~P.}\ \bibnamefont {McCulloch}}, \bibinfo
  {author} {\bibfnamefont {U.}~\bibnamefont {Schneider}}, \bibinfo {author}
  {\bibfnamefont {U.}~\bibnamefont {Schollw\"ock}}, \ and\ \bibinfo {author}
  {\bibfnamefont {F.}~\bibnamefont {Heidrich-Meisner}},\ }\bibfield  {title}
  {\enquote {\bibinfo {title} {Sudden expansion of $\mathrm{Mott}$ insulators
  in one dimension},}\ }\href {\doibase 10.1103/PhysRevB.88.235117} {\bibfield
  {journal} {\bibinfo  {journal} {Phys. Rev. B}\ }\textbf {\bibinfo {volume}
  {88}},\ \bibinfo {pages} {235117} (\bibinfo {year} {2013})}\BibitemShut
  {NoStop}%
\bibitem [{\citenamefont {Crivelli}\ \emph {et~al.}(2014)\citenamefont
  {Crivelli}, \citenamefont {Prelov\ifmmode~\check{s}\else \v{s}\fi{}ek},\ and\
  \citenamefont {Mierzejewski}}]{crivelli2014}%
  \BibitemOpen
  \bibfield  {author} {\bibinfo {author} {\bibfnamefont {D.}~\bibnamefont
  {Crivelli}}, \bibinfo {author} {\bibfnamefont {P.}~\bibnamefont
  {Prelov\ifmmode~\check{s}\else \v{s}\fi{}ek}}, \ and\ \bibinfo {author}
  {\bibfnamefont {M.}~\bibnamefont {Mierzejewski}},\ }\bibfield  {title}
  {\enquote {\bibinfo {title} {Energy and particle currents in a driven
  integrable system},}\ }\href {\doibase 10.1103/PhysRevB.90.195119} {\bibfield
   {journal} {\bibinfo  {journal} {Phys. Rev. B}\ }\textbf {\bibinfo {volume}
  {90}},\ \bibinfo {pages} {195119} (\bibinfo {year} {2014})}\BibitemShut
  {NoStop}%
\bibitem [{\citenamefont {Mendoza-Arenas}\ \emph {et~al.}(2015)\citenamefont
  {Mendoza-Arenas}, \citenamefont {Clark},\ and\ \citenamefont
  {Jaksch}}]{mendoza2015}%
  \BibitemOpen
  \bibfield  {author} {\bibinfo {author} {\bibfnamefont {J.~J.}\ \bibnamefont
  {Mendoza-Arenas}}, \bibinfo {author} {\bibfnamefont {S.~R.}\ \bibnamefont
  {Clark}}, \ and\ \bibinfo {author} {\bibfnamefont {D.}~\bibnamefont
  {Jaksch}},\ }\bibfield  {title} {\enquote {\bibinfo {title} {Coexistence of
  energy diffusion and local thermalization in nonequilibrium $\mathrm{XXZ}$
  spin chains with integrability breaking},}\ }\href {\doibase
  10.1103/PhysRevE.91.042129} {\bibfield  {journal} {\bibinfo  {journal} {Phys.
  Rev. E}\ }\textbf {\bibinfo {volume} {91}},\ \bibinfo {pages} {042129}
  (\bibinfo {year} {2015})}\BibitemShut {NoStop}%
\bibitem [{\citenamefont {Jung}\ \emph {et~al.}(2006)\citenamefont {Jung},
  \citenamefont {Helmes},\ and\ \citenamefont {Rosch}}]{rosch2006}%
  \BibitemOpen
  \bibfield  {author} {\bibinfo {author} {\bibfnamefont {P.}~\bibnamefont
  {Jung}}, \bibinfo {author} {\bibfnamefont {R.~W.}\ \bibnamefont {Helmes}}, \
  and\ \bibinfo {author} {\bibfnamefont {A.}~\bibnamefont {Rosch}},\ }\bibfield
   {title} {\enquote {\bibinfo {title} {Transport in almost integrable models:
  Perturbed $\mathrm{Heisenberg}$ chains},}\ }\href {\doibase
  10.1103/PhysRevLett.96.067202} {\bibfield  {journal} {\bibinfo  {journal}
  {Phys. Rev. Lett.}\ }\textbf {\bibinfo {volume} {96}},\ \bibinfo {pages}
  {067202} (\bibinfo {year} {2006})}\BibitemShut {NoStop}%
\bibitem [{\citenamefont {Jung}\ and\ \citenamefont {Rosch}(2007)}]{rosch2007}%
  \BibitemOpen
  \bibfield  {author} {\bibinfo {author} {\bibfnamefont {Peter}\ \bibnamefont
  {Jung}}\ and\ \bibinfo {author} {\bibfnamefont {Achim}\ \bibnamefont
  {Rosch}},\ }\bibfield  {title} {\enquote {\bibinfo {title} {Spin conductivity
  in almost integrable spin chains},}\ }\href {\doibase
  10.1103/PhysRevB.76.245108} {\bibfield  {journal} {\bibinfo  {journal} {Phys.
  Rev. B}\ }\textbf {\bibinfo {volume} {76}},\ \bibinfo {pages} {245108}
  (\bibinfo {year} {2007})}\BibitemShut {NoStop}%
\bibitem [{\citenamefont {Bamler}\ and\ \citenamefont
  {Rosch}(2015)}]{rosch1015}%
  \BibitemOpen
  \bibfield  {author} {\bibinfo {author} {\bibfnamefont {Robert}\ \bibnamefont
  {Bamler}}\ and\ \bibinfo {author} {\bibfnamefont {Achim}\ \bibnamefont
  {Rosch}},\ }\bibfield  {title} {\enquote {\bibinfo {title} {Equilibration and
  approximate conservation laws: Dipole oscillations and perfect drag of
  ultracold atoms in a harmonic trap},}\ }\href {\doibase
  10.1103/PhysRevA.91.063604} {\bibfield  {journal} {\bibinfo  {journal} {Phys.
  Rev. A}\ }\textbf {\bibinfo {volume} {91}},\ \bibinfo {pages} {063604}
  (\bibinfo {year} {2015})}\BibitemShut {NoStop}%
\bibitem [{\citenamefont {Zotos}(2004)}]{zotos2004}%
  \BibitemOpen
  \bibfield  {author} {\bibinfo {author} {\bibfnamefont {X.}~\bibnamefont
  {Zotos}},\ }\bibfield  {title} {\enquote {\bibinfo {title} {High temperature
  thermal conductivity of two-leg spin-$1/2$ ladders},}\ }\href {\doibase
  10.1103/PhysRevLett.92.067202} {\bibfield  {journal} {\bibinfo  {journal}
  {Phys. Rev. Lett.}\ }\textbf {\bibinfo {volume} {92}},\ \bibinfo {pages}
  {067202} (\bibinfo {year} {2004})}\BibitemShut {NoStop}%
\bibitem [{\citenamefont {Olshanii}(2015)}]{olshanii2015}%
  \BibitemOpen
  \bibfield  {author} {\bibinfo {author} {\bibfnamefont {Maxim}\ \bibnamefont
  {Olshanii}},\ }\bibfield  {title} {\enquote {\bibinfo {title} {Geometry of
  quantum observables and thermodynamics of small systems},}\ }\href {\doibase
  10.1103/PhysRevLett.114.060401} {\bibfield  {journal} {\bibinfo  {journal}
  {Phys. Rev. Lett.}\ }\textbf {\bibinfo {volume} {114}},\ \bibinfo {pages}
  {060401} (\bibinfo {year} {2015})}\BibitemShut {NoStop}%
\bibitem [{\citenamefont {Essler}\ \emph {et~al.}(2014)\citenamefont {Essler},
  \citenamefont {Kehrein}, \citenamefont {Manmana},\ and\ \citenamefont
  {Robinson}}]{essler2014int}%
  \BibitemOpen
  \bibfield  {author} {\bibinfo {author} {\bibfnamefont {F.~H.~L.}\
  \bibnamefont {Essler}}, \bibinfo {author} {\bibfnamefont {S.}~\bibnamefont
  {Kehrein}}, \bibinfo {author} {\bibfnamefont {S.~R.}\ \bibnamefont
  {Manmana}}, \ and\ \bibinfo {author} {\bibfnamefont {N.~J.}\ \bibnamefont
  {Robinson}},\ }\bibfield  {title} {\enquote {\bibinfo {title} {Quench
  dynamics in a model with tuneable integrability breaking},}\ }\href {\doibase
  10.1103/PhysRevB.89.165104} {\bibfield  {journal} {\bibinfo  {journal} {Phys.
  Rev. B}\ }\textbf {\bibinfo {volume} {89}},\ \bibinfo {pages} {165104}
  (\bibinfo {year} {2014})}\BibitemShut {NoStop}%
\bibitem [{\citenamefont {Huang}\ \emph {et~al.}(2013)\citenamefont {Huang},
  \citenamefont {Karrasch},\ and\ \citenamefont {Moore}}]{huang2013}%
  \BibitemOpen
  \bibfield  {author} {\bibinfo {author} {\bibfnamefont {Yichen}\ \bibnamefont
  {Huang}}, \bibinfo {author} {\bibfnamefont {C.}~\bibnamefont {Karrasch}}, \
  and\ \bibinfo {author} {\bibfnamefont {J.~E.}\ \bibnamefont {Moore}},\
  }\bibfield  {title} {\enquote {\bibinfo {title} {Scaling of electrical and
  thermal conductivities in an almost integrable chain},}\ }\href {\doibase
  10.1103/PhysRevB.88.115126} {\bibfield  {journal} {\bibinfo  {journal} {Phys.
  Rev. B}\ }\textbf {\bibinfo {volume} {88}},\ \bibinfo {pages} {115126}
  (\bibinfo {year} {2013})}\BibitemShut {NoStop}%
\bibitem [{\citenamefont {Olshanii}\ \emph {et~al.}(2012)\citenamefont
  {Olshanii}, \citenamefont {Jacobs}, \citenamefont {Rigol}, \citenamefont
  {Dunjko}, \citenamefont {Kennard},\ and\ \citenamefont
  {Yurovsky}}]{olshanii2012}%
  \BibitemOpen
  \bibfield  {author} {\bibinfo {author} {\bibfnamefont {Maxim}\ \bibnamefont
  {Olshanii}}, \bibinfo {author} {\bibfnamefont {Kurt}\ \bibnamefont {Jacobs}},
  \bibinfo {author} {\bibfnamefont {Marcos}\ \bibnamefont {Rigol}}, \bibinfo
  {author} {\bibfnamefont {Vanja}\ \bibnamefont {Dunjko}}, \bibinfo {author}
  {\bibfnamefont {Harry}\ \bibnamefont {Kennard}}, \ and\ \bibinfo {author}
  {\bibfnamefont {Vladimir~A.}\ \bibnamefont {Yurovsky}},\ }\bibfield  {title}
  {\enquote {\bibinfo {title} {An exactly solvable model for the
  integrability-chaos transition in rough quantum billiards},}\ }\href
  {\doibase 10.1038/ncomms1653} {\bibfield  {journal} {\bibinfo  {journal}
  {Nat. Commun.}\ }\textbf {\bibinfo {volume} {3}},\ \bibinfo {pages} {641}
  (\bibinfo {year} {2012})}\BibitemShut {NoStop}%
\bibitem [{\citenamefont {Prosen}(1998)}]{PRL98}%
  \BibitemOpen
  \bibfield  {author} {\bibinfo {author} {\bibfnamefont {Toma\v{z}}\
  \bibnamefont {Prosen}},\ }\bibfield  {title} {\enquote {\bibinfo {title}
  {Time evolution of a quantum many-body system: Transition from integrability
  to ergodicity in the thermodynamic limit},}\ }\href {\doibase
  10.1103/PhysRevLett.80.1808} {\bibfield  {journal} {\bibinfo  {journal}
  {Phys. Rev. Lett.}\ }\textbf {\bibinfo {volume} {80}},\ \bibinfo {pages}
  {1808--1811} (\bibinfo {year} {1998})}\BibitemShut {NoStop}%
\bibitem [{\citenamefont {Prosen}(1999)}]{PRE99}%
  \BibitemOpen
  \bibfield  {author} {\bibinfo {author} {\bibfnamefont {Toma\v{z}}\
  \bibnamefont {Prosen}},\ }\bibfield  {title} {\enquote {\bibinfo {title}
  {Ergodic properties of a generic nonintegrable quantum many-body system in
  the thermodynamic limit},}\ }\href {\doibase 10.1103/PhysRevE.60.3949}
  {\bibfield  {journal} {\bibinfo  {journal} {Phys. Rev. E}\ }\textbf {\bibinfo
  {volume} {60}},\ \bibinfo {pages} {3949--3968} (\bibinfo {year}
  {1999})}\BibitemShut {NoStop}%
\bibitem [{\citenamefont {Prosen}(2002)}]{PRE02}%
  \BibitemOpen
  \bibfield  {author} {\bibinfo {author} {\bibfnamefont {Toma\v{z}}\
  \bibnamefont {Prosen}},\ }\bibfield  {title} {\enquote {\bibinfo {title}
  {General relation between quantum ergodicity and fidelity of quantum
  dynamics},}\ }\href {\doibase 10.1103/PhysRevE.65.036208} {\bibfield
  {journal} {\bibinfo  {journal} {Phys. Rev. E}\ }\textbf {\bibinfo {volume}
  {65}},\ \bibinfo {pages} {036208} (\bibinfo {year} {2002})}\BibitemShut
  {NoStop}%
\bibitem [{\citenamefont {Berman}\ and\ \citenamefont {Izrailev}(2005)}]{FPU}%
  \BibitemOpen
  \bibfield  {author} {\bibinfo {author} {\bibfnamefont {G.~P.}\ \bibnamefont
  {Berman}}\ and\ \bibinfo {author} {\bibfnamefont {F.~M.}\ \bibnamefont
  {Izrailev}},\ }\bibfield  {title} {\enquote {\bibinfo {title} {The
  $\mathrm{Fermi-Pasta-Ulam}$ problem: Fifty years of progress},}\ }\href
  {\doibase http://dx.doi.org/10.1063/1.1855036} {\bibfield  {journal}
  {\bibinfo  {journal} {Chaos}\ }\textbf {\bibinfo {volume} {15}},\ \bibinfo
  {eid} {015104} (\bibinfo {year} {2005})}\BibitemShut {NoStop}%
\bibitem [{\citenamefont {Arnold}(1963)}]{KAM}%
  \BibitemOpen
  \bibfield  {author} {\bibinfo {author} {\bibfnamefont {V.I.}\ \bibnamefont
  {Arnold}},\ }\bibfield  {title} {\enquote {\bibinfo {title} {Proof of a
  theorem by a.n.kolmogorov on the invariance of quasi-periodic motions under
  small perturbations of the hamiltonian},}\ }\href@noop {} {\bibfield
  {journal} {\bibinfo  {journal} {Usp. Math. Nauk.}\ }\textbf {\bibinfo
  {volume} {18}},\ \bibinfo {pages} {13--40} (\bibinfo {year}
  {1963})}\BibitemShut {NoStop}%
\bibitem [{\citenamefont {Mazur}(1969)}]{mazur}%
  \BibitemOpen
  \bibfield  {author} {\bibinfo {author} {\bibfnamefont {P.}~\bibnamefont
  {Mazur}},\ }\bibfield  {title} {\enquote {\bibinfo {title} {Non-ergodicity of
  phase functions in certain systems},}\ }\href {\doibase
  http://dx.doi.org/10.1016/0031-8914(69)90185-2} {\bibfield  {journal}
  {\bibinfo  {journal} {Physica}\ }\textbf {\bibinfo {volume} {43}},\ \bibinfo
  {pages} {533 -- 545} (\bibinfo {year} {1969})}\BibitemShut {NoStop}%
\bibitem [{\citenamefont {Mori}(1965)}]{mori65}%
  \BibitemOpen
  \bibfield  {author} {\bibinfo {author} {\bibfnamefont {Hazime}\ \bibnamefont
  {Mori}},\ }\bibfield  {title} {\enquote {\bibinfo {title} {Transport,
  collective motion, and brownian motion},}\ }\href {\doibase
  10.1143/PTP.33.423} {\bibfield  {journal} {\bibinfo  {journal} {Progress of
  Theoretical Physics}\ }\textbf {\bibinfo {volume} {33}},\ \bibinfo {pages}
  {423--455} (\bibinfo {year} {1965})},\ \Eprint
  {http://arxiv.org/abs/http://ptp.oxfordjournals.org/content/33/3/423.full.pdf+html}
  {http://ptp.oxfordjournals.org/content/33/3/423.full.pdf+html} \BibitemShut
  {NoStop}%
\bibitem [{\citenamefont {Long}\ \emph {et~al.}(2003)\citenamefont {Long},
  \citenamefont {Prelov\ifmmode~\check{s}\else \v{s}\fi{}ek}, \citenamefont
  {El~Shawish}, \citenamefont {Karadamoglou},\ and\ \citenamefont
  {Zotos}}]{mclm}%
  \BibitemOpen
  \bibfield  {author} {\bibinfo {author} {\bibfnamefont {M.~W.}\ \bibnamefont
  {Long}}, \bibinfo {author} {\bibfnamefont {P.}~\bibnamefont
  {Prelov\ifmmode~\check{s}\else \v{s}\fi{}ek}}, \bibinfo {author}
  {\bibfnamefont {S.}~\bibnamefont {El~Shawish}}, \bibinfo {author}
  {\bibfnamefont {J.}~\bibnamefont {Karadamoglou}}, \ and\ \bibinfo {author}
  {\bibfnamefont {X.}~\bibnamefont {Zotos}},\ }\bibfield  {title} {\enquote
  {\bibinfo {title} {Finite-temperature dynamical correlations using the
  microcanonical ensemble and the lanczos algorithm},}\ }\href {\doibase
  10.1103/PhysRevB.68.235106} {\bibfield  {journal} {\bibinfo  {journal} {Phys.
  Rev. B}\ }\textbf {\bibinfo {volume} {68}},\ \bibinfo {pages} {235106}
  (\bibinfo {year} {2003})}\BibitemShut {NoStop}%
\end{thebibliography}%

\end{document}